\documentclass[showpacs,twocolumn,floatfix,amsmath,amssymb,superscriptaddress,prx,nopacs]{revtex4-1}
\usepackage[utf8]{inputenc}
\usepackage{epsfig}
\usepackage{graphicx}
\usepackage{longtable}
\usepackage{color}
\usepackage{dcolumn}
\usepackage{bm}
\usepackage[colorlinks=true,citecolor=magenta,linkcolor=blue]{hyperref}
\usepackage{dsfont}

\newcommand{\bea}{\begin{eqnarray}}
\newcommand{\eea}{\end{eqnarray}}
\newcommand{\bt}{\textbf}

\newcommand{\ph}{\phantom{.}}

\newcommand{\noi}{\noindent}
\newcommand{\no}{\nonumber}

\newcommand{\PK}[1]{\textcolor{black}{#1}}

\begin{document}

\title{Magnetoelectrically-Tunable Andreev-Bound-State Spectra and\\ Spin Polarization in P-Wave Josephson Junctions}
\author{Maria Teresa Mercaldo}
\affiliation{Dipartimento di Fisica ``E. R. Caianiello", Universit\`a di Salerno, IT-84084 Fisciano (SA), Italy}
\author{Panagiotis Kotetes}
\affiliation{Niels Bohr Institute, University of Copenhagen, 2100 Copenhagen, Denmark}
\affiliation{\PK{CAS Key Laboratory of Theoretical Physics, Institute of Theoretical Physics, Chinese Academy of Sciences, Beijing 100190, China}}
\author{Mario Cuoco}
\affiliation{CNR-SPIN, IT-84084 Fisciano (SA), Italy}
\affiliation{Dipartimento di Fisica ``E. R. Caianiello", Universit\`a di Salerno, IT-84084 Fisciano (SA), Italy}

\begin{abstract}
We demonstrate how the boundary-driven reconstruction of the superconducting order parameter can be employed to manipulate the zero-energy Majorana bound states (MBSs) occurring in a topological Josephson junction. We focus on an interface of two p-wave superconductors, which are described by a spin-vector order parameter $\bm{d}$. Apart from the sensitivity of $\bm{d}$ to external Zeeman/exchange fields, here, we show that the orientation of $\bm{d}$ throughout the junction can be controlled by electrically gating the weak link. The remarkable local character of this knob is a manifestation of the edge reconstruction of the order parameter, which takes place whenever different $\bm{d}$-vector configurations in each superconductor compete and are close in energy. As a consequence, the spin-dependent superconducting-phase difference across the junction is switchable from $0$ to $\pi$. Moreover, in the regime where multiple edge MBSs occur for each superconductor, the Andreev-bound-state (ABS) spectra can be twisted by the application of either a charge- or spin-phase difference across the interface, and give rise to a rich diversity of nonstandard ABS dispersions. Interestingly, some of these dispersions show band crossings protected by fermion parity, despite their $2\pi$-periodic character. These crossings additionally unlock the possibility of nontrivial topology in synthetic spaces, when considering networks of such 1D junctions. Lastly, the interface MBSs induce a distinct elecronic spin polarization near the junction, which possesses a charac\-te\-ri\-stic spatial pattern that allows the detection of MBSs using spin-polarized scanning tunneling microscopy. These findings unveil novel paths to mechanisms for ABS enginee\-ring and single-out signatures relevant for the experimental detection and manipulation of MBSs. 
\end{abstract}

\maketitle

\section{Introduction} 

The coherent control and manipulation of the electron spin are fundamental challenges in solid-state physics and the discovery of new effects and quantum materials are key prerequisite in this direction. In superconductors (SCs), Cooper pairs consist of two spin-1/2 electrons and, thus, they also support spin-1 angular momentum and orbital p-wave symmetry~\cite{Sigrist,TanakaPWave,ReadGreen,Ivanov,KitaevUnpaired,VolovikBook,Maeno2012,SatoAndo}, apart from the more conventional s-wave spin-singlet pairing. The order pa\-ra\-me\-ter in such p-wave superconductors is a complex spin vector, so-called $\bm{d}$, which ties the spin and orbital degrees of freedom together, therefore, opening perspectives for a complex response to Zeeman/ferromagnetic fields~\cite{Murakami,Dumi1,Dumi2,Wright,Mercaldo2016,Mercaldo2017,Mercaldo2018}, spin-sensitive Josephson transport~\cite{Sengupta,Yakovenko,Cuoco1,Cuoco2,BrydonJunction,Gentile} and superconduc\-ting spintro\-nics~\cite{Linder2015}. The latter two categories of phenomena can be studied by imposing a spin-phase difference $\Delta\phi_s$ across the p-wave junction, instead/on-top of a charge-phase dif\-fe\-ren\-ce $\Delta\phi_c$. The Josephson transport is mediated by the so-called Andreev bound states (ABSs), which are fermionic quasiparticle excitations appearing near the interface. The ABS spectra can be twisted by either type of superconducting-phase difference $\Delta\phi_{c,s}$, while the presence of both perturbations allows for more complex possibilities. 

Remarkably, ABSs in spin-triplet p-wave superconductors are tightly connected to Majorana quasiparticles~\cite{Majorana,Wilczek}. One-dimensional junctions of intrinsic~\cite{Sigrist,TanakaPWave,ReadGreen,Ivanov,KitaevUnpaired,VolovikBook,Maeno2012,SatoAndo} or artificial~\cite{FuKane,HasanKane,QiZhang,Alicea,CarloRev,Leijnse,KotetesClassi,Franz,Aguado,LutchynNatRevMat} p-wave superconductors harbor the so-called Majorana bound states (MBSs) which are pinned to zero energy and are charge neutral. In fact, a p-wave Josephson junc\-tion features pairs of coupled MBSs, characte\-ri\-zed by $4\pi$-periodic ener\-gy dispersions~\cite{KitaevUnpaired,Yakovenko,FuKaneJ,TanakaYokohamaNagaosa,CarloJ,KotetesJ,AliceaJ,PientkaJ,Sticlet2013,Cayao}, i.e., $\propto\cos(\Delta\phi_c/2)$. In contrast to a similar crossing appearing in d-wave superconductors~\cite{TanakaDWave,TanakaReview}, here, the linear crossing at $\Delta\phi_c=\pi$ is protected by the fermion-parity conservation~\cite{FuKaneJ,CarloJ} and reflects the exo\-tic properties of the MBSs, among which, one finds their non-Abelian exchange statistics~\cite{Ivanov}. In fact, the latter property holds promise for MBS-based topological quantum com\-pu\-ting through the non-local sto\-ra\-ge of information~\cite{Ivanov,KitaevTQC,Nayak,AliceaTQC}. This tantalizing possibility has recently sparked a plethora of predictions for MBSs in va\-rious artificial p-wave superconducting systems, inclu\-ding topological insulators proximity-coupled to SCs~\cite{FuKane}, semiconductor-superconductor hybrids~\cite{SauPRL,AliceaPRB,LutchynPRL,OregPRL}, magnetic atomic chains on superconductors~\cite{Choy,NadgPerge,Nakosai,Braunecker,Klinovaja,Vazifeh,Pientka,Ojanen1,Heimes,Brydon,Li,Heimes2,JinAn,Silas,Andolina} and o\-thers. These led to a cascade of experimental efforts veri\-fying the existence of MBSs by employing di\-ver\-se spectroscopic probes~\cite{Mourik,Yazdani1,Yacoby,ShupingLee,Ruby,Meyer,Jinfeng,Molenkamp,MT,Sven,Fabrizio,SCZhang,Yazdani2,Giazotto,HaoZhang,Attila,Wiesendanger}. Interestingly, in some of these experiments~\cite{Yazdani1,Ruby,Meyer,Jinfeng,Yazdani2,Wiesendanger}, spin-polarized scanning tunne\-ling microscopy (STM) emerged as a powerful tool to detect MBSs by means of the elecronic spin polarization they induce~\cite{Bena,IsingSpin,He,KotetesSpin}. Nonetheless, such a detection strategy has not yet been pursued to confirm intrinsic p-wave superconduc\-ti\-vi\-ty in prominent candidate materials, e.g., the Bechgaard salts~\cite{Salts1,Salts2}, lithium molybdenum purple bronze~\cite{purple}, \textcolor{black}{and more recently in the Cr-based pnictide K$_2$Cr$_3$As$_3$ \cite{Bao15,Cuono19}.}

\begin{figure*}[t!]
\centering
\includegraphics[width=0.9\textwidth]{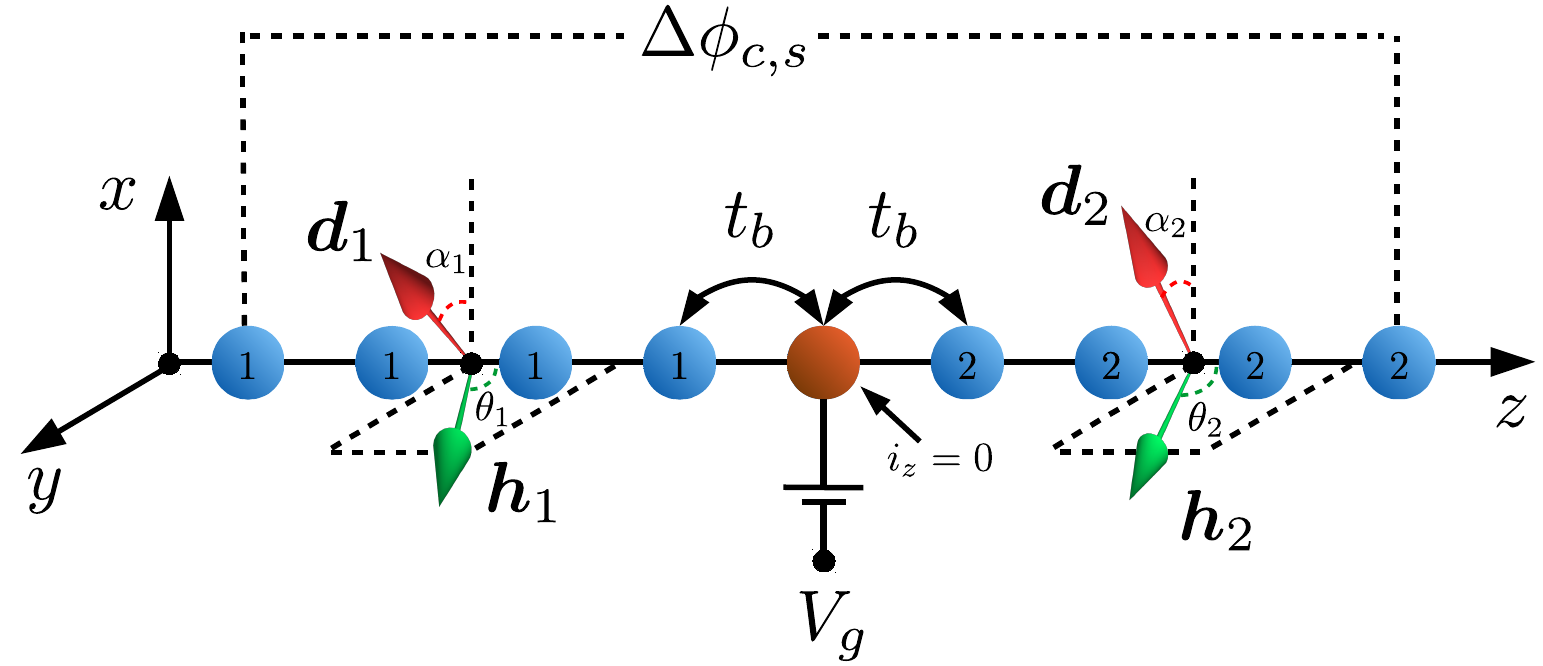}
\protect\caption{Schematic view of the heterostructure with applied fields $\bm{h}_{1,2}$, whose orientation is confined to the $yz$ plane. The fields' orientation is set by the angles $\theta_{1,2}$, which are zero for fields parallel to the $z$ direction. We depict representative $\bm{d}$-vector configurations, which are associated with the nonzero spin-triplet superconducting order parameters $\Delta_{i_z}^{\alpha\alpha}$ with $\alpha=(\uparrow,\downarrow)$. The $\bm{d}$ vectors lie in the $xy$ plane forming angles $\alpha_{1,2}$ with respect to the $x$ axis. The $i_z=0$ position labels the interface site between the two superconductors, which are labelled by $\bm{1}$ and $\bm{2}$, respectively. Each superconductor has a finite size of $N$ sites. For our numerical simulations we set $N=200$. $t_b$ is the interface charge-transfer amplitude which allows for an electronic connection between the superconductors. Gating through $V_g$ results in electrically tuning the strength of $t_b$, by suitably depleting or increasing the electronic density at the site $i_z=0$. The two superconductors feel opposite phases for both spin sectors, with the respective difference being modeled by multiplying the spin-triplet order parameters $\Delta_{i_z}^{\alpha\alpha}$ with $\alpha=(\uparrow,\downarrow)$, by the spatially dependent factors $\exp[{\rm sgn}(i_z)i \phi_{\alpha}]$. For a charge- and spin-phase difference we have $\phi_{\uparrow}= \phi_{\downarrow}=\Delta\phi_c/2$ and $\phi_{\uparrow}=-\phi_{\downarrow}=\Delta\phi_s/2$. The magnetoelectric tuning of the device leads to various Andreev-bound-state dispersions with respect to  $\Delta\phi_{c,s}$.}
\label{fig:Figure1}
\end{figure*}

In this work, we focus on control routes of MBSs and the ari\-sing ABS spectra with special emphasis on the spin-related properties, by considering Josephson junctions based on spin-triplet p-wave superconductors. Our analysis unveils a number of nonstandard ABS dispersions, which, despite their $2\pi$-periodicity with respect to the phase-difference bias, they still contain fermion-parity-protected linear band crossings. As we discuss here, and in our accompanying work of Ref.~\cite{PRL}, the presence of crossings in the ABS dispersions can form the basis for engineering nontrivial topology when considering networks of loosely-coupled 1D p-wave junctions. Notably, the nontrivial topology in the cases of interest is defined in the synthetic space spanned by the phase-difference and the wave vectors of the extra dimensions.

The wide range of ABS spectra found here, results from the competition between the energetics in the bulk and at the edge of the junction, which allow for the orien\-ta\-tion of ${\bm{d}}$ to be switched from a configuration that is coplanar to the magnetic-field's easy plane, to another, which is perpendicular to it. For a weak spin-orbit pinning, it has been shown~\cite{Wright,Mercaldo2016} that the application of a Zeeman/exchange field $\bm{h}$ induces a re-organization of the ${\bm{d}}$ vector, so that $|\bm{h}\cdot\bm{d}|$ is minimized. However, this tendency, which is promoted by the bulk contribution to the magnetic ener\-gy of the device, can be counter-balanced by boundary effects, that allow for a window of thermodynamic stability with ${\bm{d}}||\bm{h}$~\cite{Mercaldo2016}. Here, we reveal that such a topological transition can be solely controlled by electric means, i.e., by sui\-ta\-bly tuning the junction's transparency through gating at the interface. Such a local tunability is a general consequence of an edge-driven reconstruction~\cite{Mercaldo2016}, which the order parameter undergoes when dif\-fe\-rent $\bm{d}$-vector orien\-ta\-tions compete and are close in ener\-gy. 

This re-organization mechanism open perspectives for the control of the spin-phase difference across the p-wave Josephson junction. In fact, the application of Zeeman/exchange fields and interface gating potentials, can effect topological phase transitions which lead to a switching of the spin-phase difference from $0$ to $\pi$. Compared to other control mechanisms, the manipulation of the spin-triplet superconductors proposed here, has va\-rious conceptual novelties. First, it occurs by uniquely modifying a boundary condition, without changing the symmetry of the system and without going through a gap closing in the bulk. Further, it exhibits a high degree of feasibility, since the spin-phase difference control is solely based on varying the junction's transparency by tuning the strength of the charge transfer across it.  

With an eye to also motivating spin-resolved STM expe\-ri\-ments in candidate intrinsic p-wave superconductors, we additionally obtain the spatially-resolved spin-polarization of the MBSs. This exhibits a distinct spatial pattern close to the interface, which is mostly independent of the strength of the magnetoelectric drives. Moreover, different cases supporting configurations of multiple interface MBSs are demonstrated. In accordance with Ref.~\onlinecite{IsingSpin}, we show that p-wave superconductors with two MBSs per edge carry an Ising spin~\cite{IsingSpin}, which couples to Zeeman and proximity-induced exchange fields.

The paper is organized as follows. In Sec.~\ref{sec:Model} we present our model and methodology. In Sec.~\ref{sec:PhaseDiagram} we discuss the phase diagram of a magnetoelectrically-tunable p-wave junction. Specifically, we review the phenomenology of a single spin-triplet superconductor (STSC) in an external Zeeman/exchange field in Sec.~\ref{sec:SingleSTSC}, and devote Sec.~\ref{sec:DoubleSTSC} to discuss new results concerning the phase diagram of the hybrid device. Sec.~\ref{sec:ABSspectra} presents the accessible ABS spectra obtained in the various regions of the phase dia\-gram of the junction. We consider both situations of biasing the junction with a charge- and a spin-phase difference. In Sec.~\ref{sec:SpinPolarization}, we investigate the electronic spin polarization induced by MBSs, which emerge at special values of the charge- and spin-phase differences. Our spin-polarization study reveals unique properties of spin-triplet p-wave superconductors with multiple edge MBSs, which can motivate spin-resolved STM experiments. In Sec.~\ref{sec:Networks}, we extend our analysis to networks of such Josephson junctions, and focus on the type of dispersion obtained for the arising dispersive Majorana edge modes. We show that, in particular cases, such systems exhibit topologically nontrivial properties in the synthetic space composed of the phase-difference variable and the wave vector. Finally, in Sec.~\ref{sec:Conclusions}, we provide perspectives and conclusions.

\section{Model and Methodology}\label{sec:Model}

We consider a heterostructure consisting of two coupled STSCs, as shown in Fig.~\ref{fig:Figure1}, which are assumed to have been ``carved out'' of the same material. The hybrid system extends along the $z$ axis and has a length of $L=2N+1$ sites with the lattice constant set to unity. We denote the lattice sites by $i_z\in[-N,N]$. The interface of the two superconductors is located at $i_z=0$, i.e., modelled here by a single lattice point. The interface site is additionally assumed to be intrinsically non-superconducting and non-magnetic, while electrons can hop on/off it in a spin-conserving manner. Therefore, the coupling of the two superconductors, mediated by this interface link, also results to be spin-conserving. Further, each superconductor is subjected to a homogeneous Zeeman or a proximity-induced exchange field. The fields on each side of the junction generally differ on magnitude and orientation. The Bogoliubov-de Gennes (BdG) Hamiltonian in lattice space reads
\bea
{\cal H}={\cal H}_{\rm SC}+{\cal H}_{\rm SC-LINK}\,,
\eea

\noi with the term ${\cal H}_{\rm SC}$ describing the two superconducting segments defined as 
\bea
{\cal H}_{\rm SC}&=&-\sum_{\langle i_z,j_z\neq0 \rangle, \alpha} t_{i_z j_z} \big(c^{\dagger}_{i_z\alpha} c_{j_z\alpha} + \text{h.c.}\big)\no\\
&&-\sum_{i_z\neq0,\alpha,\beta}c^{\dagger}_{i_z \alpha}\big(\bm{h}_{i_z}\cdot\bm{\sigma}_{\alpha\beta}+\mu_{i_z}\delta_{\alpha\beta}\big)c_{i_z \beta}\no\\
&&+\sum_{i_z\neq0,\alpha}\left(\Delta^{\alpha\alpha}_{i_z} c^{\dagger}_{i_z\alpha} c^{\dagger}_{i_z+1\alpha} + {\text{h.c.}}\right)
\eea

\noi and the interface charge-transfer contribution given by
\bea
{\cal H}_{\rm SC-LINK}= t_b \sum_{i_z=\pm 1,\alpha}\big(c^{\dagger}_{0\alpha} c_{i_z\alpha}+{\text{h.c.}}\big)\,.
\eea

Here, $c_{i_z\alpha}$ defines the annihilation operator of an electron with spin projection $\alpha=\uparrow,\downarrow$ at the site $i_z$, and $\sigma^a$ is the Pauli matrix corresponding to the $a$-th spin direction. The hopping is nonvanishing only between the nearest-neighboring sites $\langle i_z,j_z \rangle$ with $t_{i_z j_z}=t$, while $t_b$ sets the strength of the charge transfer between the two superconductors mediated by the link. The Zeeman field is piecewise constant, ta\-king values $\bm{h}_{1,2}$ on the left and right sides of the junction, respectively. Moreover, the field is assumed to have an orientation confined to an easy plane ($yz$) throughout the device. For convenience, we parametrize the field orientation on each side using the angles $\theta_s$, i.e., $\bm{h}_s=h_s(0,\sin\theta_s,\cos\theta_s)$, with $s=1,2$. The ge\-ne\-ral bond-defined spin-triplet pairing order parameter is given by the expectation value 
\bea
\Delta^{\alpha\beta}_{i_z}=-V^{\alpha\beta}_{i_z}\langle c_{i_z+1\beta} c_{i_z\alpha}\rangle\,,
\eea 

\noi resulting from a mean-field decoupling of a suitable two-electron interaction of spin-dependent strengths $V^{\alpha\beta}_{i_z}$. The spin-triplet pairing order parameter can be expressed in the matrix form
\bea
\widehat{\Delta}_{i_z}=
\left(\begin{array}{cc}\Delta^{\uparrow\uparrow} & \Delta^{\uparrow\downarrow}\\\Delta^{\uparrow\downarrow} & \Delta^{\downarrow\downarrow}\end{array}\right)_{i_z}\equiv
\bm{d}_{i_z}\cdot\bm{\sigma}(i\sigma_y)\,,
\label{DeltaT}
\eea

\noi where we made use of the relation $\Delta^{\uparrow\downarrow}_{i_z}=\Delta^{\downarrow\uparrow}_{i_z}$, which holds by virtue of the spin-triplet character of the pai\-ring. The complex $\bm{d}$-vector components are related to the pair correlations with zero spin projection along the corresponding spin axis, and read: 
\bea
\bm{d}_{i_z}=\left(
-\frac{\Delta_{i_z}^{\uparrow\uparrow}-\Delta_{i_z}^{\downarrow\downarrow}}{2},\frac{\Delta_{i_z}^{\uparrow\uparrow}+\Delta_{i_z}^{\downarrow\downarrow}}{2i},\Delta_{i_z}^{\uparrow\downarrow}\right)\,.
\eea

In this work, the pairing interaction $V^{\alpha\beta}_{i_z}$ is assumed to be equal and spatially constant for the two superconductors, and nonzero only in the $\uparrow\uparrow$ and $\downarrow\downarrow$ channels ($V^{\uparrow\uparrow}_{i_z}=V^{\downarrow\downarrow}_{i_z}=V$). Thus, the $\bm{d}$ vector lies in the $xy$ plane, as depicted in Fig.~\ref{fig:Figure1}. We remind the reader that the pai\-ring potential is zero at the dot site $i_z=0$. Morevover, it is worth pointing out that the relative $0$ ($\pi$) spin-phase difference appearing between the $\Delta^{\uparrow\uparrow}_{i_z}$ and $\Delta^{\downarrow\downarrow}_{i_z}$ matrix ele\-ments when only the $d_y$ ($d_x$) component is present, can be pivotal for spintronic applications. In fact, we discuss in Secs.~\ref{sec:SingleSTSC} and~\ref{sec:DoubleSTSC} ways to controllably switch between ground states featuring a spin-phase difference of $0$ and $\pi$.  

In the analysis to follow, the ground state is determined for open boundary conditions by solving the BdG equation on a lattice, within an iterative self-consistent scheme of computation and by minimization of the total free energy~\cite{Cuoco1,Cuoco2}. The investigation is performed for a representative pairing-potential strength $V=2t$, lea\-ding to a $|\bm{d}|$ of the order of 0.1, and additionally assuming equal field amplitudes ($|\bm{h}_{1,2}|=h$) on both sides of the junction. The heterostructure length was set equal to $L=401$ sites for all simulations. Modifying the latter, as well as introdu\-cing small deviations from the above-mentioned parameter va\-lues, leaves the results qualitatively unchanged. 

\section{Self-consistent phase diagram of spin-triplet superconductor hybrids}\label{sec:PhaseDiagram}

Our motivation to study the particular hybrid device mainly stems from the rich response and phase diagram of a single STSC in the presence of magnetic fields~\cite{Murakami,Dumi1,Dumi2,Wright,Mercaldo2016,Mercaldo2017}, and with coexisting magnetic orders ~\cite{Mercaldo2018}. Below, we first review the results for a bulk STSC under the influence of magnetic/exhange fields, and then, we extend this study to the case of a junction consisting of two STSCs.  

\subsection{Phase diagram for a single STSC in Zeeman/exchange fields}\label{sec:SingleSTSC}

In the weak interface-coupling limit $t_b\ll t$, the phenomenology of the hybrid device can be well-understood by building upon the topological properties of the isolated superconducting segments in the presence of the corresponding Zeeman/exchange fields. Such an in-depth analysis took place by the present authors in Ref.~\onlinecite{Mercaldo2016}. We briefly review the main conclusions also here, both for reasons of completeness but, more importantly, for a smoother presentation of the new results regarding the heterostructure. 

The key result obtained for an isolated STSC in the presence of an external field, is the re-organization of the $\bm{d}$ vector due to the magnetic contribution of the MBSs to the free energy. To better understand the role of the bulk and edge degrees of freedom in the stabilization of the ground state in the presence of an applied field, let us first remind the thermodynamic and topological properties of a bulk segment of a STSC with a $\bm{d}$ vector  confined to the $xy$ plane, i.e., $\bm{d}=d(i\cos\alpha,\sin\alpha,0)$, and a field $\bm{h}=h(0,\sin\theta,\cos\theta)$ confined to the $yz$ plane. Here, $d$ is the modulus of the $\bm{d}$ vector.

In the bulk case, the type of the ground state and the resulting orientation of the $\bm{d}$ vector is solely controlled by two types of contributions to the free energy~\cite{Wright}, involving the terms $|\bm{d}\cdot\bm{h}|^2$ and $(i\bm{d}\times\bm{d}^*)\cdot\bm{h}$, respectively. The first (second) term always leads to an increase (reduction) of energy. Therefore, in order to minimize the free energy, the $\bm{d}$ vector prefers to align perpendicular to $\bm{h}$ and further develop an induced orbital magnetization $\propto i\bm{d}\times\bm{d}^*$. In conjuction with the magnetic anisotropy of the $\bm{d}$ vector assumed here, one obtains that the angle $\alpha$ is generally nonzero and is pinned by the detailed balance of the two contributions. Nevertheless, the $d_x$ component appears dominant in the weak-field limit, since it is always perpendicular to the field considered here.  

The topological properties for such a bulk system are controlled by the BdG Hamiltonian
\bea
\widehat{\cal H}_k&=&\varepsilon_k\tau_z-h\big(\sin\theta\sigma_y+\cos\theta\tau_z\sigma_z\big)\no\\
&+&2d\sin k\big(\cos\alpha\tau_y\sigma_z-\sin\alpha\tau_y\big)\label{eq:BdG}
\eea

\noi defined for the spinor 
\bea
C_k^{\dag}=(c_{k\uparrow}^{\dag},c_{k\downarrow}^{\dag},c_{-k\uparrow},c_{-k\downarrow})\,.
\eea

\noi Each Hamiltonian term is represented using Kronecker products of the Nambu $\bm{\tau}$ and spin $\bm{\sigma}$ Pauli matrices, together with the corresponding unit matrices which we omit throughout this manuscript for simplicity. In addition, we introduced the dispersion $\varepsilon_k=-2t\cos k-\mu$. 

The BdG Hamiltonian above resides in the BDI symmetry class exhi\-bi\-ting chiral, time-reversal and charge-conjugation symmetries~\cite{Altland,KitaevClassi,Ryu,TeoKane,Shiozaki,Chiu} with corre\-spon\-ding ope\-ra\-tors: $\Pi=\tau_x\sigma_z$, $\Theta=\sigma_z{\cal K}$ and $\Xi=\tau_x{\cal K}$. Here, ${\cal K}$ stands for complex conjugation. We insist that the emer\-ging time-reversal symmetry does not lead to a Kramers degeneracy, because it satisfies $\Theta^2=I$, with $I$ the identity operator. The particular symmetry class allows for $0,1,2$ MBSs per edge for this Hamiltonian when open boun\-dary conditions are considered, and under the condition that edge-driven re-organization effects are not taken into account~\cite{Mercaldo2016,Mercaldo2017}. The MBS state vectors are in this case eigenstates of the chiral-symmetry operator $\Pi$, i.e.: 
\begin{align}
\big|\Pi=+1\big>=\big\{\left|\tau_x=+1;\sigma_z=+1\right>,\left|\tau_x=-1;\sigma_z=-1\right>\big\},\no\\
\big|\Pi=-1\big>=\big\{\left|\tau_x=+1;\sigma_z=-1\right>,\left|\tau_x=-1;\sigma_z=+1\right>\big\}.
\end{align}

\begin{figure*}[t!]
\centering
\includegraphics[width=0.4\textwidth]{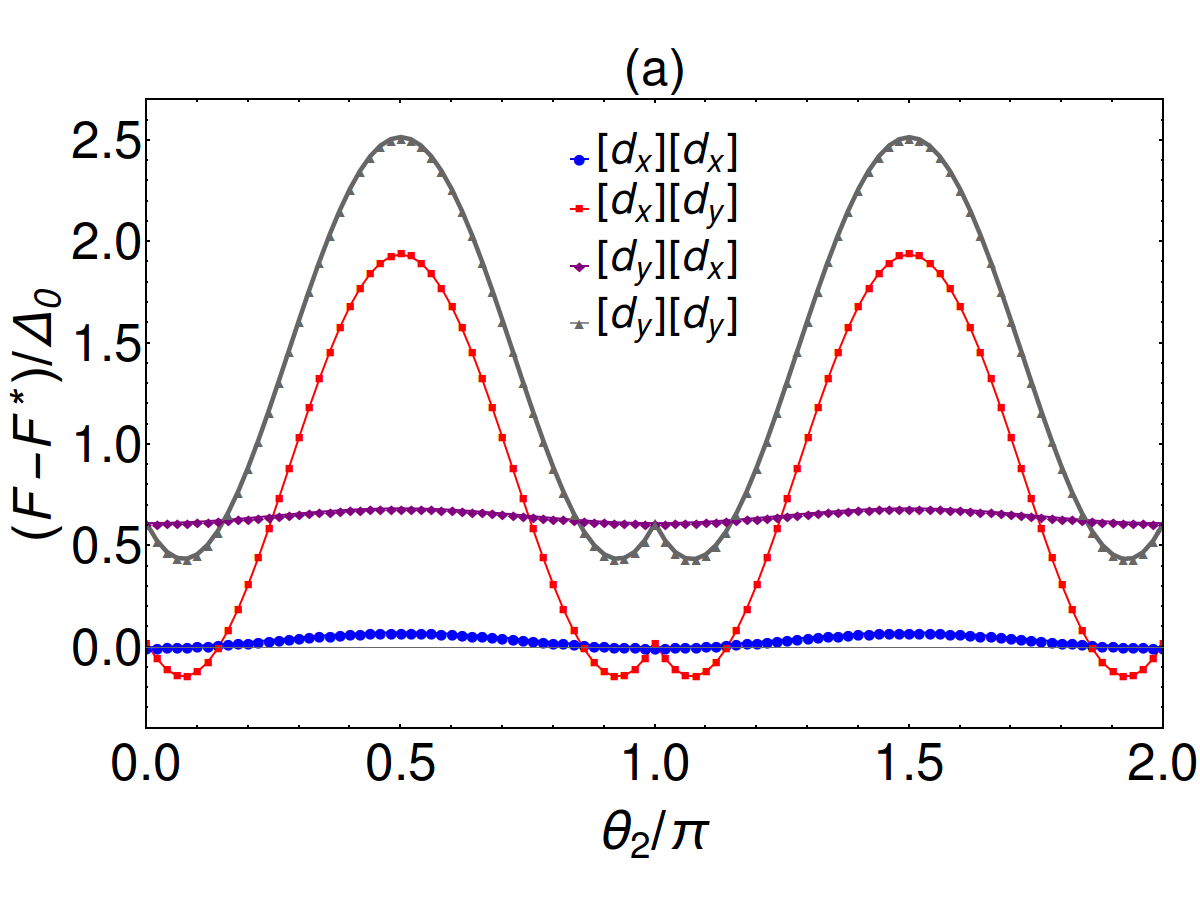}\hspace{0.4cm}
\includegraphics[width=0.28\textwidth]{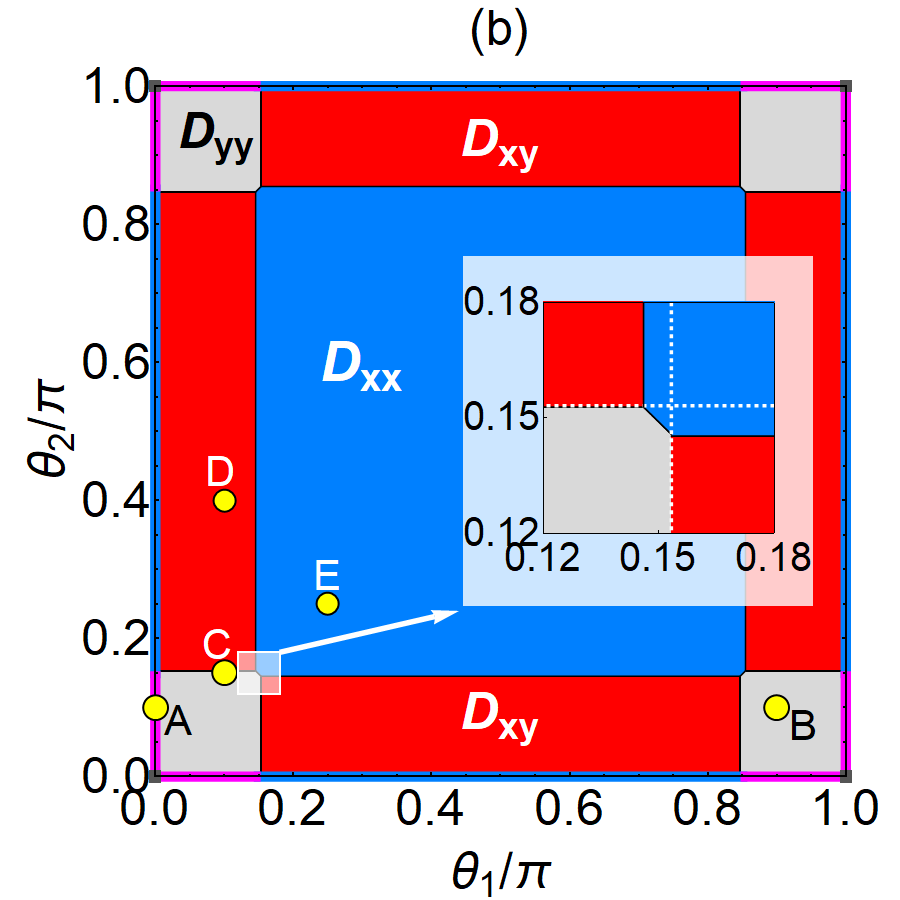}\hspace{-0.1cm}
\includegraphics[width=0.28\textwidth]{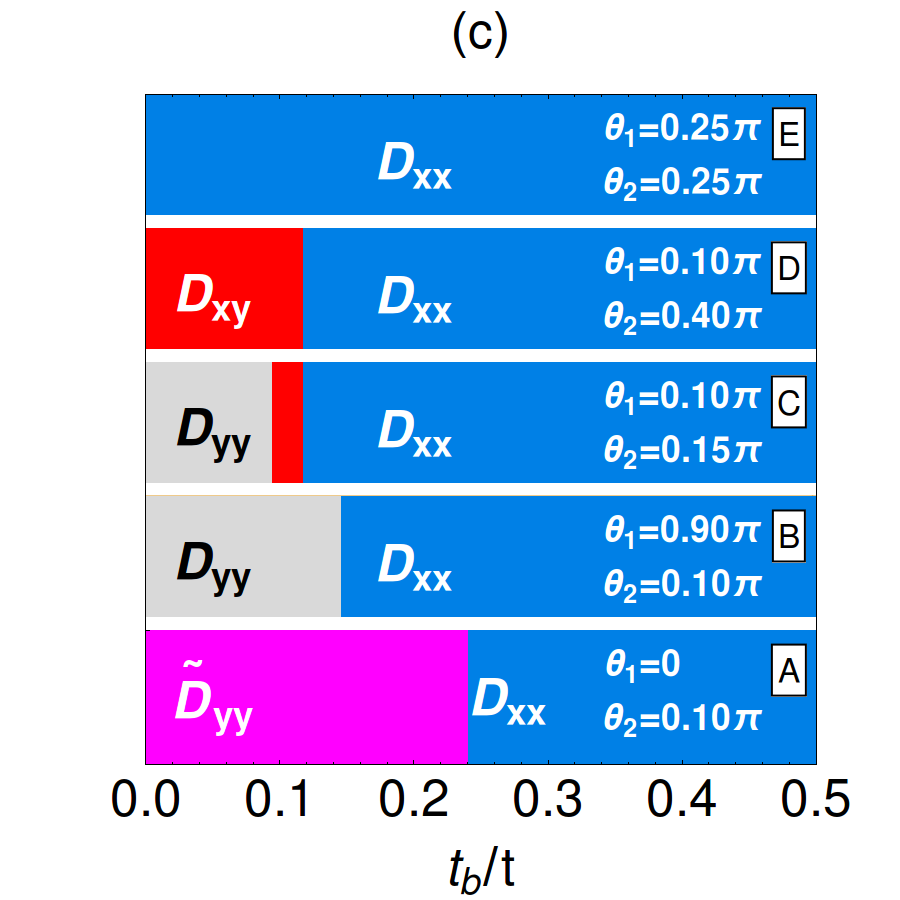}\\
\protect\caption{(a) Comparison of the computed free energy $F$ for various $\bm{d}$-vector orientations, as a function of $\theta_2$ at a fixed orientation of $\bm{h}_1$. The notation $[d_a][d_b]$ indicates a configuration with a dominant $d_a$ component in SC$_1$ and $d_b$ in SC$_2$ ($a,b=x,y$). $F^*$ is an energy offset introduced for graphical convenience \big[i.e. $F^*=F_{[d_x][d_x]}(\theta_2=0)$\big], $\Delta_{0}$ is the SC gap at zero field, and the parameters are $t=1$, $h_1=h_2=0.05t$, $\mu=1.0t$, $t_b=0.05 t$ and $\theta_1=\pi/4$. (b) Phase diagram in terms of the field-rotation angles for a representative set of parameters as in (a). Different colors refer to inequivalent configurations of the $\bm{d}$ vector, specifically $D_{ab}$ corresponds to the region in which the most stable configuration has a dominant $d_a$ component in one SC and $d_b$ in the other ($a,b=x,y$). In the inset, an enlargement of the shaded square is shown, where the white lines are the phase boundaries between the $D_{ab}$ zones, when the two SCs are disconnencted ($t_b=0$). The various regions have different topological behavior: phases in the $D_{xx}$ (blue) region harbor $4$ MBSs at the interface when the phase difference $\Delta\phi_c$ across the junction is $\pi$. Phases in the $D_{xy}$ (red) area support 2 MBSs at the interface for $\Delta\phi_c=0$ and $\pi$. The $D_{yy}$ (grey) regions define topologically-trivial phases, in the sense that MBSs can never become re-entrant for any value of $\Delta\phi_{c}$. In addition, the $\tilde{D}_{yy}$ (magenta) slivers, take place when one of the fields is oriented along the $z$ axis. Remarkably, while also in this case the order-parameter configuration is $d_y$-$d_y$, the topological behavior is radically different from the $D_{yy}$ configuration. Note, that, the yellow points indicate the positions of the phase diagram where the analysis refers to a variation of the hopping amplitude $t_b$ at the interface, shown in panel (c).}
\label{fig:Figure2}
\end{figure*}

In the limiting case $d_y=0$ with $\alpha=0,\pi$ ($d_x=0$ with $\alpha=\pm\pi/2$), we obtain two (zero) MBSs per edge. Thus, when the $d_x$ component is zero, the system becomes tri\-vial. This is quite remar\-kable, since in the absence of the field the presence of either $d_x$ or $d_y$ yields two MBSs per edge. For a ground state with only the $d_x$ component nonzero, and zero field, the MBS eigenstates are ge\-ne\-ral\-ly linear combinations of the ones of the BdG Hamiltonian of Eq.~\eqref{eq:BdG} and, thus, eigenstates of $\Pi$. However, the MBS-state vector structure is markedly different for a ground state with only the $d_y$ component nonzero. In this case, the zero-field MBS solutions on the two edges are eigenstates of the chiral-symmetry operator $\tilde{\Pi}=\tau_x$, and read:
\begin{align}
\big|\tilde{\Pi}=+1\big>=\big\{\left|\tau_x=+1;\sigma_z=+1\right>,\left|\tau_x=+1;\sigma_z=-1\right>\big\},\no\\
\big|\tilde{\Pi}=-1\big>=\big\{\left|\tau_x=-1;\sigma_z=+1\right>,\left|\tau_x=-1;\sigma_z=-1\right>\big\}.
\end{align}

\noi This notable difference of the MBS state-vector structure accounts for the different behavior of a ground state with only a nonzero $d_x$ or $d_y$ component, when the field is added. Since the $d_x$ case is adiabatically connected to the general case of Eq.~\eqref{eq:BdG}, we expect the two MBSs per edge to persist in the weak-field limit. Instead, in the $d_y$ case, the pair of MBSs at a given edge persists if the applied field is oriented along the $z$ axis, while they become hybridized into nonzero energy ABSs for a field oriented along the $y$ axis.

The above discontinuous topological behavior, depen\-ding on whether $d_y$ is the only nonzero $\bm{d}$-vector component, has drastic consequences on deciding the ground state of the open system in the small $\theta$ angle and weak-field limit. When only the $d_y$ component is present, the pair of MBSs at each edge hybridize into nonzero ABSs and this conversion process lowers the free energy. Therefore, there exists a narrow window in which the local hybridization of MBS pairs dominates the cost of simultaneously increasing $|\bm{d}\cdot\bm{h}|$ and decreasing $|\bm{d}\times\bm{d}^*|$, with both tendencies imposed by the bulk degrees of freedom. Conclusively, an isolated and finite-sized STSC, under the conditions assumed here, is dictated by a nonzero $d_y$ component for small field angle and amplitude, while, away from this limit, both $d_{x,y}$ components are generally present with their relative angle $\alpha$ mainly controlled by the bulk contribution to the free energy.  

\subsection{Phase diagram for a STSC junction in Zeeman/exchange fields}\label{sec:DoubleSTSC}

In this section we investigate the phase diagram of the hybrid device of the two indirectly-coupled STSCs, in terms of the applied-field orientation and the strength of the charge transfer across the interface. We aim at exploring the effects of the interface coupling on the ground state of the system and the emergence of MBSs. We exa\-mi\-ne both the weak and strong limits for the interface coupling strength $t_b$, thus, allowing us to interpolate from a tunneling-like to the high-transparency regime. Our analysis mainly focuses on the most-interesting weak-field regime, since this is the one allowing for multiple MBSs per edge. 

Notably, if an increase of the amplitude of the exchange field is allowed, then an odd number of MBSs can also appear at the interface. For this circumstance, there is always a single interface MBS which remains at zero energy for any type and value of the applied phase dif\-fe\-ren\-ces across the junction~\cite{Sticlet2013}. For the applications we have in mind, this case appears less attractive compared to the one with an even number of MBSs per edge. Therefore, we do not consider it in the remainder.

\subsubsection{Phase diagram as a function of the Zeeman fields' orientations for weak interface coupling}

One of the principal goals of the present study is to investigate how Zeeman or exchange fields, present on both sides of the junction, control the relative $\bm{d}$-vector orientations of the two STSC segments. In the weak interface-coupling limit $t_b\ll t$, the resulting ground state can be practically inferred by viewing the two superconducting segments as disconnected. Then, the edge reconstruction of the $\bm{d}$ vector along the junction follows from the discussion of Sec.~\ref{sec:SingleSTSC}. In this case, the interface coupling can be considered a weak perturbation, mainly influencing the MBSs which are located near the interface. Given these arguments, the superconducting con\-fi\-gu\-ra\-tion on each side is either $(d_x\ll d_y)$ or $(d_y\ll d_x)$, when weak external fields are con\-si\-de\-red. Therefore, in the remainder, the $\bm{d}$-vector con\-fi\-gu\-ra\-tion of the hybrid device will be marked by the do\-mi\-nant $d_x$ or $d_y$ component of each side, i.e., $[d_a][d_b]\rightarrow D_{ab}$ with $a,b=x,y$. Hence, in this weak-field limit, we can identify four possible he\-te\-ro\-struc\-tu\-re configurations. 
 
We start by numerically exploring the phase diagram in terms of varying the field orientations, while pre\-ser\-ving the field's magnitude constant throughout the two superconducting segments. The spatial profile of the order parameter is obtained via an iterative self-consistent scheme of computation until the desired accuracy is achieved throughout the entire heterostructure. Apart from being confined in the $xy$ plane, the $\bm{d}$ vector is otherwise unconstrained during this search. 

Representative profiles of the free energy associated with the $\bm{d}$ vector solutions are reported in Fig.~\ref{fig:Figure2}(a). As an example, we show that for a field orientation $\theta_1=\pi/4$ of $\bm{h}_1$, the variation of $\theta_2$ can drive different $\bm{d}$-vector flop transitions from $d_x$ to $d_y$. See also Fig.~\ref{fig:Figure2}(b) for a tho\-rough phase-diagram exploration in terms of the $(\theta_1,\theta_2)$ parameter space for $t_b=0.05t$. Note, that, identical copies of this phase diagram are obtained in the remaining three quadrants of $(\theta_1,\theta_2)$ space after extending $\theta_{1,2}\in[0,2\pi)$. In Figs.~\ref{fig:Figure2}(a) and (b), we observe that the ground state of the heterostructure can change from $D_{xy}$ to $D_{xx}$ when the orientation angle $\theta_2$, of the exchange field $\bm{h}_2$, de\-via\-tes sufficiently from $0$ or $\pi$. 

As one would also expect from the results of a single STSC interfaced with vacuum, the application of the field on each side of the superconductor is able to pin the $\bm{d}$ vector along the $x$ or $y$ direction in spin space. In addition, according to the intui\-ti\-ve arguments discussed in the previous paragraph, we indeed find that it is this ``local'' re-organization mechanism which shapes the entire form of the phase diagram. As a matter of fact, when scanning the parameter space, we find that the $D_{xx}$ region dominates the phase diagram, especially for orientations of the exchange fields which are far from being parallel to the $z$ direction, i.e., $\theta_{1,2}\sim 0,\pi$. This is because, in this case, the $h_y$ component is substantial and leads to an energy cost (through $|\bm{d}\cdot\bm{h}|^2$) which exceeds the gain from the hybridization of the edge MBSs accessible via the stabilization of the $(d_x=0,d_y\neq0)$ ground state. Conversely, when $\theta_{1,2}$ are near $0$ or $\pi$, we obtain a $D_{yy}$ ground state with all the interface MBSs hybridized, even for $t_b=0$. However, if one of the two angles is equal to $0$ or $\pi$, then one obtains the here so-denoted $\tilde{D}_{yy}$ ground state, which can harbor two MBSs per edge for the superconductor under the influence of the field parallel to the $z$ axis. The $D_{xy}$ phase is stabilized if one field is oriented close to $0$ or $\pi$, i.e., parallel to $z$, while the other one is away from this range of orientations. 

\subsubsection{Representative phase diagrams as a function of the interface-coupling strength}

As discussed in the previous section, in the weak interface-coupling strength limit it is essentially the influen\-ce of the field on each side of the heterostructure which decides on the spatial profile of the $\bm{d}$ vector. Nonetheless, the impact of a stronger interface coupling is expected to be relevant near the $(\theta_1,\theta_2)$ pa\-ra\-me\-ter regions of the phase diagram in which the four distinct device ground states meet. This happens, for instance, when the applied fields have an orientation given by the angles $\theta_1=\theta_2 \sim 0.15 \pi$. To resolve the role of the crosstalk between the two superconducting segments, we here explore the modification of the phase diagram upon a variation of the charge-transfer strength. 

Such an exploration unveils a general trend regarding the transitions occurring upon increasing $t_b$, i.e., when moving from a regime of bad $t_b<0.1$ to good $t_b \sim 0.5$ electronic interface-mat\-ching of the two superconductors. According to this trend, independently of the ground-state configuration that becomes stabilized by the external fields in the weak interface-coupling limit, the increase of $t_b$ leads to a $D_{xx}$ phase. This is also illustrated in Fig.~\ref{fig:Figure2}(c). Remarkably, the control of the barrier strength opens the path to a local manipulation of the ground state and the topological properties of the heterostructure, solely through the tunability of the interface energy contribution without involving any bulk energy-spectrum gap closings. These results also indicate, that, rendering the interface further transparent, reduces the available boundary regions and the possible energy gain from the hybridization of the interface MBSs and, thus, favors the $D_{xx}$ configuration which is promoted by the bulk degrees of freedom.

Finally, we investigate the special situation of field orien\-ta\-tions along the $z$ direction, i.e., $\theta_{1,2}= 0,\pi$. In this case, each isolated SC can reside in either the $d_x$ or $d_y$ ground states, since they are degenerate. In fact, in either configuration, each SC harbors two MBSs per edge, which are protected by a unitary symmetry that allows disconnecting the two spin sectors. Therefore, before contact, all the $D_{ab}$ ground states with $a,b=x,y$ are degenerate. However, switching on $t_b$ renders the $D_{ab}$ with $a\neq b$ always energetically unfavorable.

\section{Andreev-bound-state Spectra}\label{sec:ABSspectra}
    
In this paragraph we shed light on the topological pro\-per\-ties of the hybrid device by considering a charge ($\Delta\phi_c$) or a spin ($\Delta\phi_s$) superconduc\-ting-phase difference across the junction. The study of the low-energy dispersions with respect to these two types of phase differences, reveals important information regarding the hybridization of the underlying MBSs, their possible zero-energy-pinning restoration, as well as the arising Josephson currents. 

To evaluate the consequences of a phase gradient across the junction, we introduce both a charge- and a spin-phase difference between the left and right sides of the he\-te\-ro\-struc\-tu\-re. See also Fig.~\ref{fig:Figure1}. This is effected by multiplying the spin-triplet pairing order parameters $\Delta_{i_z}^{\alpha\alpha}$, with $\alpha=(\uparrow,\downarrow)$, by the spatially dependent factors $\exp[{\rm sgn}(i_z)i \phi_{\alpha}]$. In this manner, the two STSCs feel opposite phases. Hence, for the case of a conventional charge Josephson configuration, we consider that both spin channels are twisted by the same phase factor, i.e., $\phi_{\uparrow}= \phi_{\downarrow}=\Delta\phi_c/2$. On the other hand, to examine the spin component of the supercurrent flowing across the heterostructure, one has to consider that $\phi_{\uparrow}=-\phi_{\downarrow}=\Delta\phi_s/2$.

Depending on the field angles $\theta_{1,2}$ and the $\bm{d}_{1,2}$ vectors stabilized, the STSCs comprising the junction can reside in a topologically non-trivial regime before contact and, thus, allow for edge MBSs. As discussed in Sec.~\ref{sec:SingleSTSC}, in the general case, each STSC features a BDI symmetry classification when isolated, therefore, giving rise to an integer number of edge MBSs protected by chiral symmetry. On top of that, one encounters excep\-tional si\-tua\-tions in which either all the MBSs become hybridized or, if present, they are protected by a unitary, instead of a chiral, symmetry. In all cases, the interface MBSs can in principle hybridize and combine into nonzero-energy fermionic ABSs, or, persist being pinned at zero ener\-gy if additional symmetry constraints apply. 

\begin{figure*}[t!]
\centering
\includegraphics[width=0.97\textwidth]{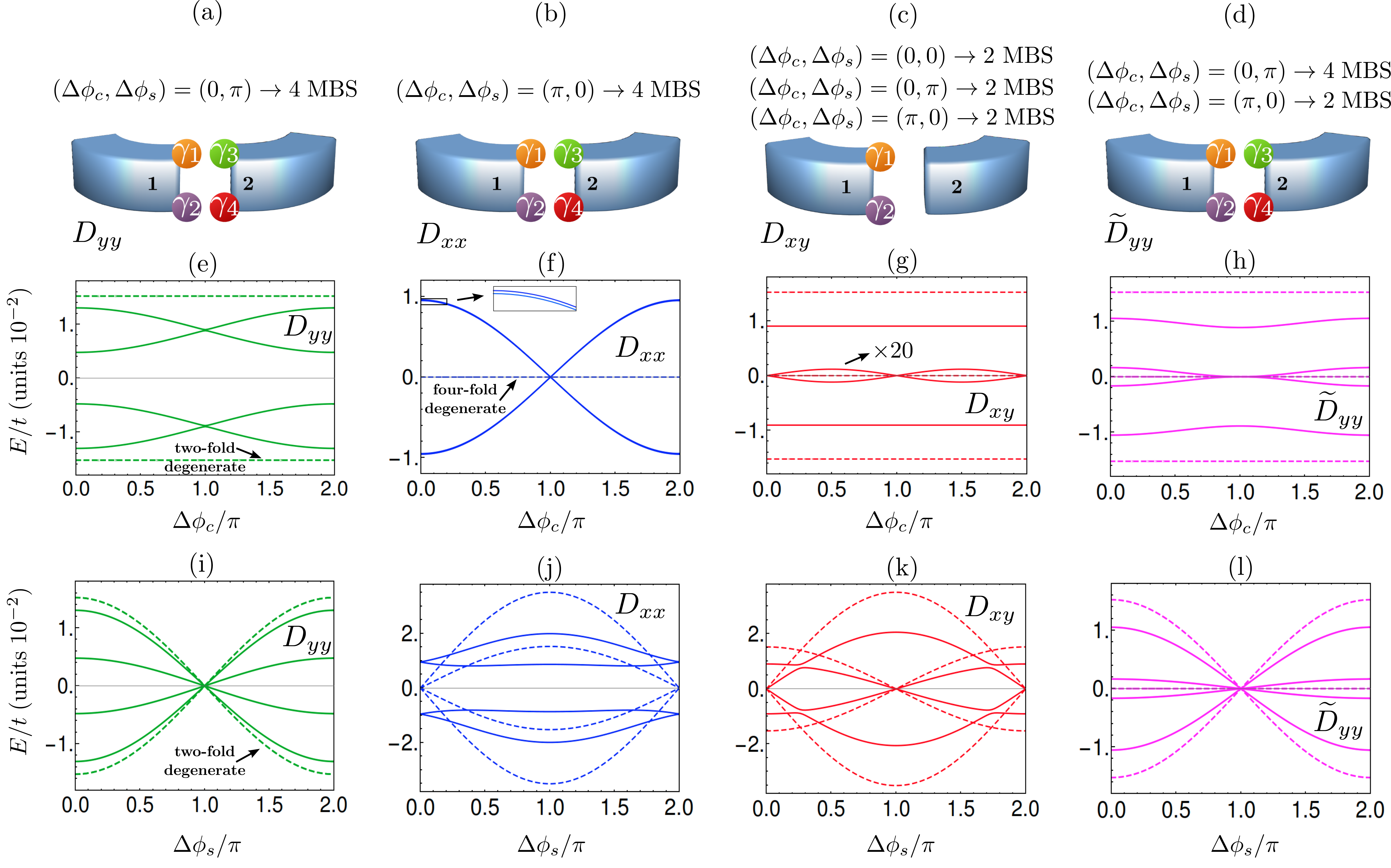}
\protect\caption{(a)-(d) Sketch of the MBSs occurring at the interface of the STSC heterostructure for the $D_{yy}$, $D_{xx}$, $D_{xy}$ and $\tilde{D}_{yy}$ regions of the phase diagram shown in Fig.~\ref{fig:Figure2}. In (e)-(h) and (i)-(l) we depict the low-energy spectra as a function of the charge- and spin-phase differences $\Delta\phi_c$ and $\Delta\phi_s$, imposed between the two STSCs. In each plot, eight energy levels are shown, accounting for both the four nearby interface (solid lines) and the four far-away bound states (dashed lines). The parameters employed are as follows: for $D_{yy}$ we used $\theta_1=\theta_2=0.1\pi$ and $t_b=0.10t$; for $D_{xx}$ we have $\theta_1=0.1\pi$, $\theta_2=0.25\pi$ and $t_b=0.15t$; for $D_{xy}$ we chose $\theta_1=0.1\pi$, $\theta_2=0.25\pi$ and $t_b=0.05t$;  at last, for $\tilde{D}_{yy}$ we considered $\theta_1=0$, $\theta_2=0.1\pi$ and $t_b=0.10t$. For all the plots we assumed $t=1$, $h_1=h_2=0.05t$ and $\mu=t$. The $D_{yy}$ state in (e) and (i) exhibits always fully-gapped ABS spectra, except for $\Delta\phi_s=\pi$, which leads to the crossing of all the eight modes at zero-energy. The ABS spectra for the $D_{xx}$ configuration are shown in (f) and (j). In the presence of $\Delta\phi_c$ there are always four degenerate zero-energy solutions which correspond to the MBSs far from the interface (dashed lines), and this condition holds for $\Delta\phi_s=0$. At $\Delta\phi_c=\pi$, the MBSs at the interface of the heterostructures are restored, thus, resulting to the eight zero-energy solutions shown in (f). As shown in (g), the ABS spectra for the $D_{xy}$ case which is dictated by very different $\bm{d}$-vectors on the two sides of the junction, are practically nondispersive upon applying a charge-phase drop across the interface. Nonetheless, as shown in (k), the dispersion of these bands becomes enhanced in the presence of a spin-phase difference $\Delta\phi_s$. In this case, one finds that four degenerate zero-energy modes are restored for the high-symmetry values $\Delta\phi_s=0,\pi$. In panel (h) we focus on the $\tilde{D}_{yy}$ configuration, in which, the MBS at the interface are split by all the possible values for the charge-phase difference, except for $\Delta\phi_c=\pi$. As portrayed in (l), the application of a spin-phase difference allows for four MBSs to emerge at the interface when $\Delta\phi_s=\pi$.}
\label{fig:Figure3}
\end{figure*}

Here, having considered only short-ranged hopping amplitudes and weak interface-coupling and field strengths, a maximal number of two MBSs per edge can be obtained. See Ref.~\cite{Mercaldo2016} for other possibilities. Given these conditions, we are left with only two possible $\bm{d}$-vector configurations which yield MBSs, i.e., either with the $d_x$ component necessarily nonzero for any orientation of the field in the $yz$ plane, or, for only the $d_y$ component nonzero and the respective field exact\-ly aligned along the $z$ axis. Taking into account the phase diagram in Fig.~\ref{fig:Figure2}, we depict in Figs.~\ref{fig:Figure3}(a)-(d) the MBSs obtained in each device ground state for special values of the phase dif\-fe\-ren\-ces $\Delta\phi_{c,s}$. We note, that, for a charge-phase difference interface MBSs can reappear only for the $D_{xx}$, $D_{xy}$ and $\tilde{D}_{yy}$ device ground states. Instead, interface MBSs can only re-emerge for the $D_{yy}$, $D_{xy}$ and $\tilde{D}_{yy}$ device ground states when applying a spin-phase difference. Below we provide a theoretical framework to understand these fin\-dings.

Since we are interested in limits of weak Zeeman/exchange fields and low transparency for the junction, we may infer the qualitative and, up to a certain extend, the quantitative properties of the hybrid device by projecting the full Hamiltonian describing the system onto the MBS subspace. For the analytical discussion to follow, it is convenient to integrate out the link degrees of freedom at $i_z=0$, and obtain the effective interface coupling between the two superconductors
\begin{align}
{\cal H}_{\rm SC-SC}=\sum_{i_z,j_z>0}\sum_{\alpha=\uparrow,\downarrow}{\rm T}_{i_z+j_z}\left(c_{-i_z\alpha}^{\dag}c_{j_z\alpha}+{\rm h.c.}\right)\,.\label{eq:SC-SC-0}
\end{align} 

\noi Here, ${\rm T}_{i_z+j_z}$ define the matrix elements for the effective electron hopping between the two SCs, which depend only on the sites' distance, the details of the weak link, and the barrier strength $t_b$. The next step is to gauge away the piecewise spatially-varying phases from the order parameters $\Delta_{i_z}^{\alpha\alpha}$ with $\alpha=\uparrow,\downarrow$. This is effected via the spin-dependent gauge transformation:
\bea
c_{i_z\alpha}\mapsto e^{{\rm sgn}(i_z)i \phi_{\alpha}/2}c_{i_z\alpha}\,,
\eea

\noi which transfers the phase differences to the interface-coupling Hamiltonian of Eq.~\eqref{eq:SC-SC-0}, which in the new frame reads:
\begin{align}
{\cal H}_{\rm SC-SC}^{\Delta\phi_{c,s}}=\sum_{i_z,j_z>0}{\rm T}_{i_z+j_z}\left(\bm{c}_{-i_z}^{\dag}e^{i(\Delta\phi_c+\Delta\phi_s\sigma_z)/2}\bm{c}_{j_z}+{\rm h.c.}\right)
\label{eq:SC-SC}
\end{align} 

\noi where we introduced the spinor $\bm{c}_{i_z}^{\dag}=(c_{i_z\uparrow}^{\dag}\ph c_{i_z\downarrow}^{\dag})$. Notably, the spin-part of this gauge transformation further modifies the Zeeman/exchange fields felt by the two SCs, thus, yielding in the new frame:
\bea
\bm{h}_1&\mapsto&(+h_{1y}\sin(\Delta\phi_s/2),h_{1y}\cos(\Delta\phi_s/2),h_{1z})\,,\label{eq:Gaugeh1}\\
\bm{h}_2&\mapsto&(-h_{2y}\sin(\Delta\phi_s/2),h_{2y}\cos(\Delta\phi_s/2),h_{2z})\,.\label{eq:Gaugeh2}
\eea

\noi The above, implies that in this new basis, the spin-phase bias induces nonzero $x$ components for the fields felt by the two SCs. Noteworthy, when the magnetic field has all its components nonzero, it violates the chiral symmetry protecting the pairs of edge MBSs for the ground state of a single STSC with $d_{x,y}\neq0$. In this scenario, the MBSs at a given side of the interface hybridize for all values of $\Delta\phi_s$, since the magnetic field always possesses a component in the $\bm{d}$-vector's plane. However, for a single STSC with only $d_y$ nonzero, the two vectors $\bm{h}$ and $\bm{d}$ become orthogonal for $\Delta\phi_s=\pi$, thus, zero-energy MBSs emerge. 

\textcolor{black}{We additionally remark, that, in the event of a net equilibrium current bia\-sing the junction, the phase $\Delta\phi_c$ of Eq.~\ref{eq:SC-SC} has to be shifted by an amount proportional to the line integral of the vector potential $A_z$. This follows from the expression for the spin-resolved ($\alpha=\uparrow,\downarrow$) current density $\bm{J}_{\alpha}(\bm{r})=\bm{\nabla}\phi_{\alpha}+(2e/\hbar)\bm{A}(\bm{r})$, which holds when the contribution of the quasiparticles can be neglected. Nevertheless, since in the present section we are interested in 1D junctions which are solely phase biased, such a phase shift is not present. Evenmore, for the open 1D geometry exa\-mi\-ned here, the fact that the electrons couple only to the $z$ component of the vector potential, further implies that there is no orbital coupling to the applied magnetic field. However, the orbital coupling becomes relevant in quasi 1D setups (see also Sec.~\ref{sec:Networks}).}

In the following analysis we take into account the mo\-di\-fi\-cation of the interface-coupling Hamiltonian, as well as the rotation of the fieds, and project them onto the MBS subspace to analyze the ABS spectra for the four possible device ground states.

\subsection{$\mathbf{D_{yy}}$ ground state}

Based on the preceding discussions, it is straightforward to understand the phenomenology of the $D_{yy}$ device ground state, with the main results depicted in Fig.~\ref{fig:Figure3}(a). In Figs.~\ref{fig:Figure3}(e) and (i) we show the ABS spectra for a nonzero charge- and spin-phase dif\-fe\-ren\-ce, respectively. In the first case, all interface MBSs are gapped out due to the nonzero $y$ component of the magnetic field, which breaks the unitary symmetry protecting the pairs of MBSs appearing before contact at zero field. In contrast, when a spin-phase bias is imposed, the unitary symmetry becomes reinforced for $\pi$, thus, restoring the protected pairs of MBSs of the decoupled STSCs. In fact, the MBSs at $\pi$ remain present even after the two STSCs come in contact, since also the Majorana-Josephson couplings vanish at the same spin-phase value. Consequently, one obtains fermion-parity-protected li\-near crossings at $\Delta\phi_s=\pi$.

\subsection{$\mathbf{D_{xx}}$ ground state}

In this ground state, the left (right) supercon\-ducting segment harbors two interface MBSs, with correspon\-ding operators $\gamma_{1;\Pi_1;\nu}$ ($\gamma_{2;\Pi_2,\nu}$). Here, $\nu=\pm1$ labels operators posses\-sing the same chi\-ra\-li\-ty, which is an eigenvalue of the chiral-symmetry operator $\Pi=\tau_x\sigma_z$. Notably, in the weak-field limit, the chirality of the MBSs is essentially decided by the sign of the $d_x$ component and, thus, can be inverted by a $\pi$-rotation of the field in the $yz$ plane. Given a specific chi\-ra\-li\-ty value, the respective MBS eigenstates can be found using the results of Sec.~\ref{sec:SingleSTSC}. There exist two scenarios, linked to the two possibilities: $\Pi_1=\Pi_2$ or $\Pi_1=-\Pi_2$.

We first examine the case of $\Pi_1=-\Pi_2$ and assume, without loss of generality, that $\Pi_1=+1$. At this stage, we project Eq.~\eqref{eq:SC-SC} onto the MBS subspace, by approxima\-ting the electronic operators as follows: 
\bea
\bm{c}_{i_z<0}&\approx& 
f_{i_z}\left(\begin{array}{c}1\\0\end{array}\right)\gamma_{1;\Pi_1=+1;+1}\ph+g_{i_z}\left(\begin{array}{c}0\\i\end{array}\right)\gamma_{1;\Pi_1=+1;-1}
\no\\
\bm{c}_{i_z>0}&\approx& 
\tilde{f}_{i_z}\left(\begin{array}{c}i\\0\end{array}\right)\gamma_{2;\Pi_2=-1;+1}\ph+\tilde{g}_{i_z}\left(\begin{array}{c}0\\1\end{array}\right)\gamma_{2;\Pi_2=-1;-1}\no\\
\eea

\noi with $f_{i_z}$, $g_{i_z}$, $\tilde{f}_{i_z}$ and $\tilde{g}_{i_z}$ appropriate functions decaying away from the interface. After introducing the above expressions in Eq.~\eqref{eq:SC-SC}, we find the low-energy Hamiltonian coupling MBSs belonging to different TSCs
\begin{align}
{\cal H}_{\rm MBS}^{\rm inter}=\sum_{\nu=\pm1}t_{\nu}\cos\left(\frac{\Delta\phi_c+\nu \Delta\phi_s}{2}\right)i\gamma_{1;\Pi_1;\nu}\gamma_{2;-\Pi_1;\nu}
\end{align}

\noi with $t_{\pm1}$ appropriate coupling matrix elements proportional to $t_b$. For $\Delta\phi_s=0$, it is only the above Hamiltonian that couples the interface MBSs and leads to two ABS branches. These ABS spectra exhibit the usual $4\pi$-periodic Majorana-Josephson dependence, with linear cros\-sings at $\Delta\phi_c=\pi$, which are now protected by chiral symmetry, instead of fermion-parity. This result is numerically confirmed, as it becomes evident from Fig.~\ref{fig:Figure3}(f). For $\Delta\phi_c=0$, all the interface MBSs are split, while the MBSs away from the interface stay at zero energy for all values of $\Delta\phi_c$. Instead, when a spin-phase difference is imposed, the ABS spectrum originating from interface MBSs is fully gapped, as also illustrated in Fig.~\ref{fig:Figure3}(j). As mentioned earlier, this is a consequence of the violation of chiral symmetry for a nonzero $\Delta\phi_s$, which pairs-up MBSs at all edges. Note that the MBSs away from the interface remain at zero energy for $\Delta\phi_{c,s}=0$, due to the preservation of chiral symmetry and the vanishing of the inter-STSC-coupling matrix elements ${\rm T}_{i_z+j_z}$ near the very-left and very-right edges.

We conclude this paragraph with the scenario $\Pi_1=\Pi_2$, in which all the interface MBSs share the same chirality. As a result, the inter-STSC couplings are now proportional to $\sin\left[(\Delta\phi_c+\nu \Delta\phi_s)/2\right]$, instead of $\cos\left[(\Delta\phi_c+\nu\Delta\phi_s)/2\right]$, with $\nu=\pm1$. This shift does not lead to any qualitative difference compared to the already-examined case of opposite chiralities.

\subsection{$\mathbf{D_{xy}}$ ground state}

We continue with the $D_{xy}$ ground state. In this case, the left segment yields two zero-energy interface MBSs $\gamma_{1;\Pi_1;\pm1}$, while the right segment yields two already-split MBSs. Since here we are interested in the weak-field limit, we may build our analysis for the right segment upon the MBSs which are obtained in zero field. To correctly describe the right segment we then need to expli\-cit\-ly add the MBS-hybridization term induced by the $h_y$ field component. This hybridization term is already present for zero phase differences. Within this approach, the MBSs for the right segment are associated with the two operators $\gamma_{2;\tilde{\Pi}_2,\pm1}$. Notably, the pairs of MBSs on the two sides of the interface are protected by different chiral-symmetry ope\-ra\-tors. This mismatch plays a crucial role for the perio\-di\-city of the energy dispersions with respect to the phase differences. 

To elaborate on the periodicity properties of the phase-difference-twisted ABS spectra for such a hybrid-device ground state, we consider a particular example where $\Pi_1=-\tilde{\Pi}_2=+1$. Thus, without loss of gene\-ra\-li\-ty, we appro\-xi\-ma\-te the electronic operators in the follow sense:
\bea
\bm{c}_{i_z<0}&\approx& f_{i_z}\left(\begin{array}{c}1\\0\end{array}\right)\gamma_{1;\Pi_1=+1;+1}+g_{i_z}\left(\begin{array}{c}0\\i\end{array}\right)\gamma_{1;\Pi_1=+1;-1}
\no\\
\bm{c}_{i_z>0}&\approx& 
\tilde{f}_{i_z}\left(\begin{array}{c}i\\0\end{array}\right)\gamma_{2;\tilde{\Pi}_2=-1;+1}+\tilde{g}_{i_z}\left(\begin{array}{c}0\\i\end{array}\right)\gamma_{2;\tilde{\Pi}_2=-1;-1}\,.\no\\
\eea

\noi Projection of the coupling Hamiltonian in Eq.~\eqref{eq:SC-SC} and taking into account the symmetry-breaking effects of the field for the two STSCs, yield the MBS interface couplings
\bea
{\cal H}_{\rm MBS}
&=&t_+\cos\left(\frac{\Delta\phi_c+\Delta\phi_s}{2}\right)i\gamma_{1;\Pi_1=+1;+1}\gamma_{2;\tilde{\Pi}_2=-1;+1}\no\\
&+&t_-\sin\left(\frac{\Delta\phi_c-\Delta\phi_s}{2}\right)i\gamma_{1;\Pi_1=+1;-1}\gamma_{2;\tilde{\Pi}_2=-1;-1}\no\\
&+&E_{h_{1y}}\sin(\Delta\phi_s/2)i \gamma_{1;\Pi_1=+1;+1}\gamma_{1;\Pi_1=+1;-1}\no\\
&-&E_{h_{2y}}\cos(\Delta\phi_s/2)i\gamma_{2;\tilde{\Pi}_2=-1;+1}\gamma_{2;\tilde{\Pi}_2=-1;-1}\,,
\eea

\noi with $t_{\pm}\propto t_b$, $E_{h_{1y}}\propto h_{1y}$ and $E_{h_{2y}}\propto h_{2y}$ suitable ener\-gy scales. We remark that also the MBSs away from the interface feel the two last field-related terms of the Hamiltonian above. Therefore, the two pairs of the far-away edge MBS always become split for $\Delta\phi_{s}\neq0,\pi$.

From the above, we find that when $\Delta\phi_s=0$, two far-away and two interface MBSs become pinned to zero ener\-gy for a charge-phase difference of $0$ or $\pi$. The former (latter) are located only (predominantly) at the edges of the STSC on the left hand side. Remarkably, the situation is reversed for $\Delta\phi_c=0,\pi$ and a spin-phase difference of $\Delta\phi_s=\pi$, with the far-away (interface) MBSs appea\-ring only (mainly) at the edges of the STSC on the right hand side. See also Figs.~\ref{fig:Figure3}(g) and~(k). In fact, this result is independent of the precise values of the chiralities of the interface MBSs. 

We remark, that, for both cases of applying a charge- or a spin-phase difference, we find an ABS branch which, despite exhibiting a $2\pi$-phase periodicity, it contains two fermion-parity protected linear crossings. For $t_{\pm}\ll E_{h_{2y}}$ ($t_{\pm}\ll E_{h_{1y}}$) we can integrate out the ABS fermion associated with the pair of split MBSs on the right (left) segment, and obtain an effective coupling for the two interface MBSs of the left (right) segment alone. This is only possible for $\Delta\phi_s$ away from $\pi$ ($0$). For an illustration, we set $\Delta\phi_s=0$ and obtain the following effective low-energy Hamiltonian:
\begin{align}
{\cal H}_{\rm low-en}\approx\frac{t_+t_-}{2E_{h_{2y}}}\sin\Delta\phi_c\ph i\gamma_{1;\Pi_1=+1;+1}\gamma_{1;\Pi_1=+1;-1}\,.
\end{align}

\noi The resulting two ABS dispersions account for the sinusoidal-like energy branches shown in Fig.~\ref{fig:Figure3}(g). Si\-mi\-lar arguments allow us to obtain a low-energy sinusoidal ABS branch dispersing with $\Delta\phi_s$, for $\Delta\phi_c=0,\pi$. This result is shown in Fig.~\ref{fig:Figure3}(k) and can be seen as the result of an intrinsic nonzero $\pi$ spin-phase difference across the interface in the $D_{xy}$, which is rooted in the $\bm{d}$-vector mismatch. Even more, the presence of a sinusoidal-like ABS branch leads to a spontaneous spin-Josephson current.

\subsection{$\mathbf{\tilde{D}_{yy}}$ ground state}

The final type of device ground state to be discussed is the case in which both segments feature a $d_y$ con\-fi\-gu\-ra\-tion. However, on one side the field is aligned along the $z$ axis, while on the other, the field is generally orien\-ted in the $yz$ plane. As in the previous paragraph, the weak-field limit considered here allows us to use the zero-field case as a starting point. At zero field we obtain two pairs of symmetry-proteted MBSs per interface side, by means of a chiral symmetry which is present. This is generated by the operator $\tilde{\Pi}=\tau_x$ for both STSCs. When the abovementioned field configuration is switched on, the pair of MBSs on one side become split due to the $h_y$ component of the field. Thus, one can distinguish two possibilities depending on whether the chiralities $\tilde{\Pi}_{1,2}$ of the interface MBSs are the same or opposite. 

We first examine the case of opposite chiralities, and for the demonstration we assume that $\tilde{\Pi}=+1$ for the STSC on the left hand side, which is additionally assumed here to be under the influence of the field which is parallel to the $z$ axis. Given these assumptions, we have the MBS-projected electron operators: 
\bea
\bm{c}_{i_z<0}&\approx& f_{i_z}\left(\begin{array}{c}1\\0\end{array}\right)\gamma_{1;\tilde{\Pi}_1=+1;+1}+
g_{i_z}\left(\begin{array}{c}0\\1\end{array}\right)\gamma_{1;\tilde{\Pi}_1=+1;-1}
\no\\
\bm{c}_{i_z>0}&\approx& 
\tilde{f}_{i_z}\left(\begin{array}{c}i\\0\end{array}\right)\gamma_{2;\tilde{\Pi}_2=-1;+1}+
\tilde{g}_{i_z}\left(\begin{array}{c}0\\i\end{array}\right)\gamma_{2;\tilde{\Pi}_2=-1;-1}\,.\no\\
\eea

\noi Projection of the coupling Hamiltonian and the effects of the field on the right segment, yields the MBS couplings
\bea
{\cal H}_{\rm MBS}
&=&t_+\cos\left(\frac{\Delta\phi_c+\Delta\phi_s}{2}\right) i\gamma_{1;\tilde{\Pi}_1=+1;+1}\gamma_{2;\tilde{\Pi}_2=-1;+1}\no\\
&+&t_-\cos\left(\frac{\Delta\phi_c-\Delta\phi_s}{2}\right) i\gamma_{1;\tilde{\Pi}_1=+1;-1}\gamma_{2;\tilde{\Pi}_2=-1;-1}\no\\
&-&E_{h_{2y}}\cos(\Delta\phi_s/2)i\gamma_{2;\tilde{\Pi}_2=-1;+1}\gamma_{2;\tilde{\Pi}_2=-1;-1}\,.
\eea

\noi We observe that for $\Delta\phi_s=\pi$ the above Hamiltonian yields two $4\pi$-periodic Majorana-Josephson dispersions with respect to $\Delta\phi_c$ which, however, are proportional to $\sin(\Delta\phi_c/2)$ instead of $\cos(\Delta\phi_c/2)$. Instead, if $\Delta\phi_s\neq\pi$, the ABS spectrum is $2\pi$-periodic in $\Delta\phi_c$. If $\Delta\phi_c=0$, we find a linear crossing at $\Delta\phi_s=\pi$ for all branches, since in this case chiral symmetry on the right hand side segment is restored and, at the same time, all the inter-STSC MBS couplings vanish. See Figs.~\ref{fig:Figure3}(h) and~(l) for the numerical confirmation of these results, obtained using the fully-self-consistent formalism.

Similar to the $D_{xy}$ case, we can find an approximate expression for the ABS spectra as a function of $\Delta\phi_c$ when $t_{\pm}\ll E_{h_{2y}}$ and $\Delta\phi_s$ sufficiently away from $\pi$. To connect to the results of Fig.~\ref{fig:Figure3} we set $\Delta\phi_s=0$. By integrating out the MBSs on the right segment we find an effective coupling for the two MBSs of the left segment, which reads
\begin{align}
\bar{{\cal H}}_{\rm low-en}\approx\frac{t_+t_-}{2E_{h_{2y}}}\big(1+\cos\Delta\phi_c\big)i\gamma_{1;\tilde{\Pi}_1=+1;+1}\gamma_{1;\tilde{\Pi}_1=+1;-1}\,.
\end{align}

\noi The resulting positive and negative ABS branches agree with the ones obtained numerically and are shown in Fig.~\ref{fig:Figure3}(h). It becomes evident from inspecting the above expression, that, there exists a quadratic band crossing at $\Delta\phi_c=\pi$.

\section{Majorana bound states and spin polarization}\label{sec:SpinPolarization}

Recent groundbreaking experiments~\cite{Yazdani1,Ruby,Meyer,Jinfeng,Yazdani2,Wiesendanger} have detected the presence of MBSs in artificial to\-po\-lo\-gical super\-conductors which rely on magnetic chains deposited on top of a superconducting substrate. In these experiments, MBSs appear in the effectively-spinless limit and their observed fingerprints are associated with the nontrivial spin content of the corresponding MBS wavefunctions~\cite{Bena}, rather than a true spin degree of freedom. Essentially, the spin-content of the MBS wavefunction enters in the matrix elements describing the coupling of the MBS to the STM-tip electrons and, this, leads to a spin-selective differential conductance~\cite{He,KotetesSpin}. This situation maps to the case of an initially spinful STSC, such as in Eq.~\eqref{eq:BdG}, when the field is sufficiently strong to completely deplete one of the two spin bands. In this effectively-spinless limit, the STSC harbors a single MBS per edge. Nevertheless, both intrinsic and artificial systems can exhibit multiple MBSs per edge, despite the fact that the spin degree of freedom is practically quenched. 

As we have discussed throughout this work, such a possibility arises in the presence of a symmetry, e.g., of the chiral type. For instance, such a scenario is unlocked for the system of Eq.~\eqref{eq:BdG} if we consider the spinless limit, but assume a $\bm{d}$ vector with a $k$ structure of the form $\sin(nk)$, with $n\in\mathbb{Z}$. In the case $n=2$, two MBSs become possible on the same edge, which remain uncoupled by virtue of the orthogonal spatial distributions of the correspon\-ding wavefunctions~\cite{Heimes2}. Thus, here, it is a sublattice degree of freedom which leads to two MBSs per edge. In fact, if we also replace the $\cos k$ nearest-neighbor hopping term of Eq.~\eqref{eq:BdG} with its next-nearest-neighbor analog, i.e., $\cos(2k)$, we obtain two decoupled inter\-pe\-ne\-tra\-ting sublattices, each one supporting a single MBS per edge. The possibility of multiple chiral-symmetry-protected MBSs has been already studied in the STSC context in Refs.~\onlinecite{Dumi1,Dumi2,Mercaldo2016,Mercaldo2017}. Notably, a similar si\-tua\-tion takes place in topological magnetic chains as, also there, it has been shown~\cite{Heimes2,JinAn,Silas,Andolina,KotetesSpin} that multiple MBS can emerge by virtue of a sublattice/chiral symmetry.   

In the present study, however, we deal with a radically different physical situation, since the STSCs in discussion are studied in the weak-field limit. In fact, the intrinsic topological nature of a STSC allows it to harbor two MBSs per edge without the requirement of a magnetic perturbation. For the $k$ structure of the pai\-ring term considered in Eq.~\eqref{eq:BdG}, one obtains two MBSs per edge by virtue of a unitary symmetry, which essentially reflects the time-reversal symmetry of the system at zero fields. Therefore, the electron spin is an active degree of freedom in our case and can leave a unique imprint on the spin-resolved STM, which goes beyond the spin-selectivity effects discussed in Refs.~\onlinecite{He,KotetesSpin}. In fact, the authors of Ref.~\cite{IsingSpin}, have shown that a Kramers pair of MBSs is characterized by an Ising spin, i.e., the spin-density ope\-ra\-tor is nonzero only along a specific direction in spin space. In stark contrast, the spin-density operator in the case of multiple MBSs protected by a sublattice symmetry is identically zero~\cite{Bena}. Nonetheless, the \textit{electron} spin-density induced by MBSs is nonzero even for effectively-spinless topological superconductors, and this exactly what the spin-resolved experiments of Refs.~\onlinecite{Yazdani1,Ruby,Meyer,Jinfeng,Yazdani2,Wiesendanger} have measured so far.

In our present analysis we move towards both directions. We first demonstrate the emergence of an Ising spin for the uncoupled MBSs in the STSC junction, which appear for special values of $\Delta\phi_{c,s}$, and additionally compute the electronic component of the spin density induced by these MBSs. In the following, we focus on the $D_{xx}$ configuration, while a similar qualitative behavior is obtained for the $D_{xy}$ and $\tilde{D}_{yy}$ device ground states, which are briefly discussed in Appendix~\ref{sec:Appendix}.  

\begin{figure*}[t!]
\includegraphics[width=0.6\textwidth]{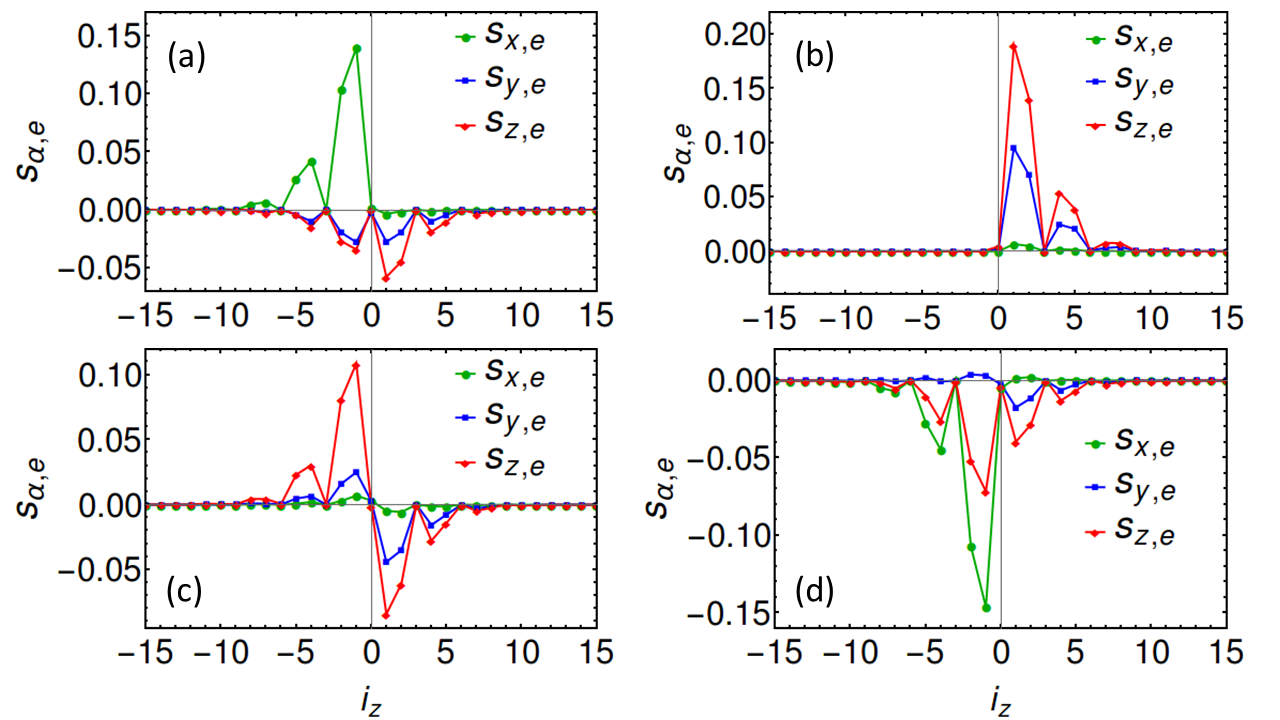}
\includegraphics[width=0.35\textwidth]{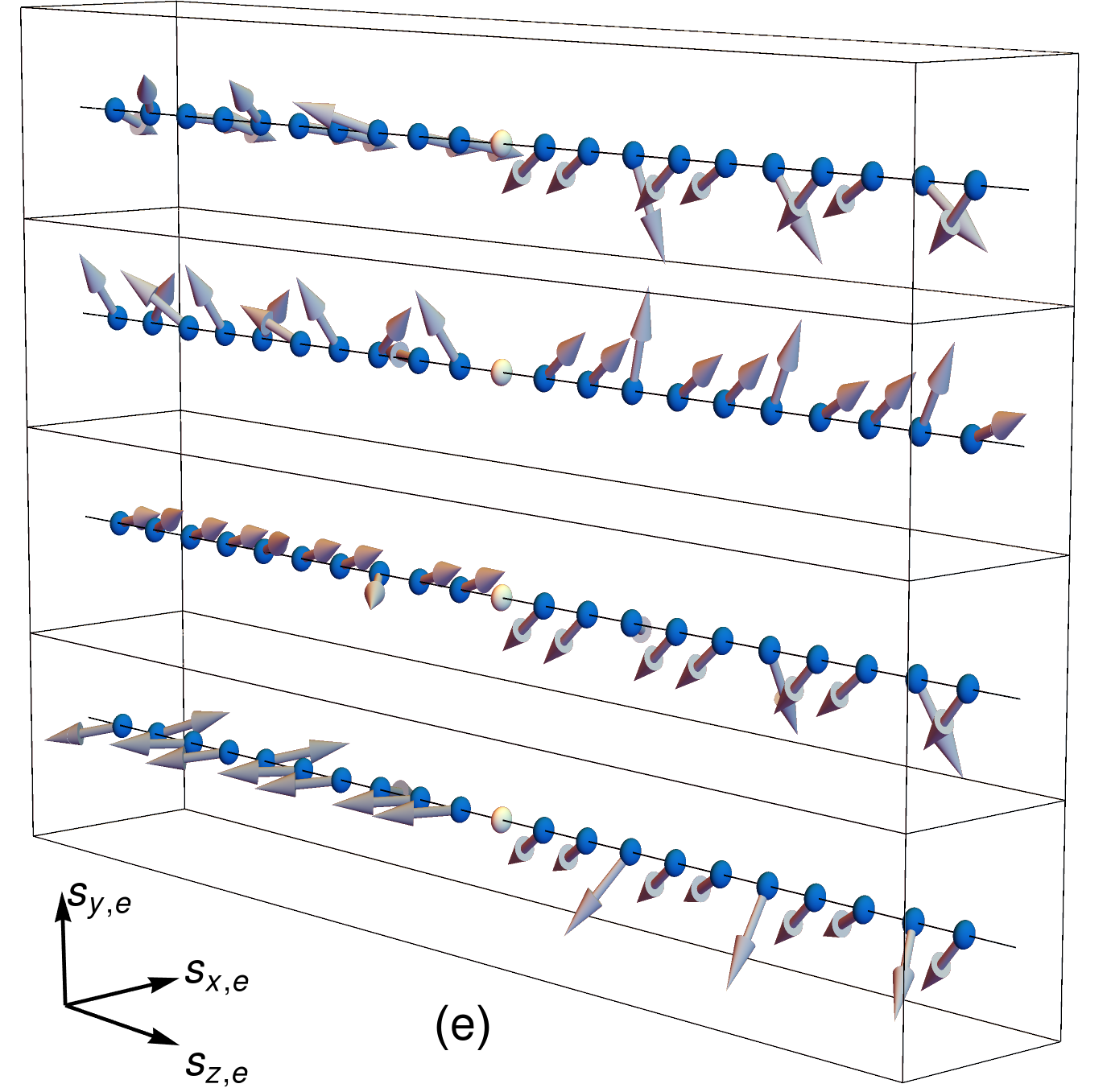}
\protect\caption{(a)-(d) Electron component of the spin-polarization for each MBS for a representative case of the $D_{xx}$ topological phase, with 4 MBSs occurring close to the interface assuming that the phase difference is $\Delta\phi_c=\pi$ and $\Delta\phi_s=0$. The parameters are: $\theta_1=0.10\pi$, $\theta_2=0.25 \pi$ and $t_b=0.15t$. In the right panel (e) we report the atomically-resolved three-dimensional texture of the MBS-induced electronic spin-polarization close to the interface site, which is indicated with a golden point. \textcolor{black}{We point out that the MBSs depicted in (a)-(d) and (b)-(c) belong to the same chirality subspace, and their respective wavefunctions are linear combinations of the ones appearing in Eqs.~\eqref{eq:SpinMBS1}-\eqref{eq:SpinMBS2} and/or \eqref{eq:SpinMBS3}-\eqref{eq:SpinMBS4}. This becomes manifest in the behaviour of the corresponding $s_{x,e}$ components, which for a given pair, have the same amplitude but opposite sign.}}
\label{fig:Figure4}
\end{figure*}

\subsection{Ising spin-density operator}

In this section, we focus on the Ising structure of the spin density emerging in the $D_{xx}$ ground state for $\Delta\phi_c=\pi$ and $\Delta\phi_s=0$. In this case, two pairs of uncoupled MBSs appear at the junction's interface as a result of the restoration of chiral symmetry. Note that the chiral symmetry in the present case can be connected to a time-reversal, rather than a sublattice, symmetry in the zero-field limit, thus, allowing the spin to be a relevant degree of freedom. 

We begin our analysis by considering the structure of the MBS wavefunctions, using the real-space spinor basis 
\bea
C_{i_z}^{\dag}=\left(c_{i_z\uparrow}^{\dag},\, c_{i_z\downarrow}^{\dag},\, c_{i_z\uparrow},\, c_{i_z\downarrow}\right)\,.
\eea

\noi Since the BdG Hamiltonian has both the charge-conjugation $\Xi=\tau_x{\cal K}$ and chiral $\Pi=\tau_x\sigma_z$ symmetries, the two MBSs wavefunctions ($\nu=\pm1$) for a given chi\-ra\-li\-ty ($\Pi=\pm 1$) have the following general form:
\bea
\Phi_{{\rm MBS},\Pi=+1,+1}=\sum_{i_z} \frac{g_{+1}(i_z)}{\sqrt{2}}\left(\begin{array}{c}  \sin(\beta_{i_z}) \\ - i \cos(\beta_{i_z}) \\
\sin(\beta_{i_z})\\ i \cos(\beta_{i_z}) \end{array}\right),\quad\label{eq:SpinMBS1}\\
\Phi_{{\rm MBS},\Pi=+1,-1}=\sum_{i_z} \frac{g_{-1}(i_z)}{\sqrt{2}}\left(\begin{array}{c} \cos(\beta_{i_z}) \\  i \sin(\beta_{i_z}) \\ \cos(\beta_{i_z})\\ -i \sin(\beta_{i_z}) \end{array}\right),\quad\label{eq:SpinMBS2}\\
\Phi_{{\rm MBS},\Pi=-1,+1}=\sum_{i_z} \frac{g_{+1}(i_z)}{\sqrt{2}}\left(\begin{array}{c} - i \sin(\beta_{i_z}) \\  - \cos(\beta_{i_z}) \\
i \sin(\beta_{i_z})\\ - \cos(\beta_{i_z}) \end{array}\right),\quad\label{eq:SpinMBS3}\\
\Phi_{{\rm MBS},\Pi=-1,-1}=\sum_{i_z}\frac{g_{-1}(i_z)}{\sqrt{2}} \left(\begin{array}{c} -i \cos(\beta_{i_z}) \\  \sin(\beta_{i_z}) \\ i  \cos(\beta_{i_z})\\ \sin(\beta_{i_z}) \end{array}\right).\quad\label{eq:SpinMBS4}
\eea

\noi with $g_{\pm1}(i_z)$ being functions that take into account the spatial distribution of the MBS wavefuntions near the interface of the junction, while the angle $\beta_{i_z}$ sets the local electron- and hole-spin orientations. Hence, taking into account that the spin operator is expressed as:
\bea
\bm{s}=m_s\left(\tau_z\sigma_x,\sigma_y,\tau_z\sigma_z\right)
\eea

\noi with $m_s=\hbar/2$, one can immediately verify that the expectation values satisfy $\big<\Phi_{{\rm MBS},\Pi,\nu}\big|\bm{s}\big|\Phi_{{\rm MBS},\Pi,\nu}\big>=\bm{0}$, independently of the values for $\nu$ and $\Pi$. On the other hand, the $x$ component of the spin operator has a nonvanishing amplitude when considering the mixing term $\left<s_x\right>_{i_z}=\big<\Phi_{{\rm MBS},\Pi,\nu}\big|s_x\big|\Phi_{{\rm MBS},\Pi,-\nu}\big>\sim i g_{\nu}(i_z)^{*} g_{-\nu}(i_z)$. Then, one generally finds that, for a given chirality subspace, the expectation value of the spin polarization including both the electron and hole parts, is identically zero for the components which are coplanar with the magnetic field (i.e. $yz$ plane), while it is not vanishing for the projections along the direction perpendicular to the magnetic easy plane. Thus, as also demonstrated in Ref.~\onlinecite{IsingSpin}, each pair of MBSs can exhibit an Ising-type spin polarization with an orientation which is perpendicular to the plane of the applied magnetic field. We remark, that, the averaged spin polarization over all the MBSs is vanishing. 

\subsection{MBS-induced electronic spin polarization}

While the interference of the electron and hole part generally builds up a spin polarization along a par\-ti\-cu\-lar spin orientation, the investigation of the projected electron or hole components of the MBSs unveils a much richer structure in both the spin and spatial distribution. But more importantly, it is the electron component of the spin density which is accessed at the atomic scale through STM. In this case, the measured spin-resolved electron current is probing the local electron component of the BdG eigenstate, at the given ener\-gy associated with the applied voltage. Therefore, here we are interested in zero-bias spin-resolved STM. 

The inspection of the electron component of the MBS wavefunctions shows a distinct spatial and orien\-ta\-tion dependence. The first observation is that each MBS wavefunction exhibits a spin profile in space with local components which are nonvanishing along all the symmetry directions. See Figs.~\ref{fig:Figure4}(a)-(d). This can be tracked by using again the general expression of the MBS wavefunctions and obtain only the spin ope\-ra\-tor projected onto the electron component, i.e., $\bm{s}_e={\cal P}_e\bm{s}$ with ${\cal P}_e=(\mathds{1}+\tau_z)/2$ the corresponding projector. One now finds that all the components of $\bm{s}_e$ are now non\-va\-ni\-shing. For instance, the local $y$ and $z$ spin components in the space spanned by $\{\Phi_{{\rm MBS},\Pi,\nu}$,$\Phi_{{\rm MBS},\Pi,-\nu}\}$ have both diagonal and off-diagonal terms, e.g., for the $z$ electronic component this is given by $\left<s_{z,e} \right>_{\nu=1;\nu=1}=-\left<s_{z,e} \right>_{\nu=-1;\nu=-1}=-m_s\cos(2\beta)/2$ and $\left< s_{z,e} \right>_{\nu;-\nu}=m_s\sin(2\beta)/2$. On the other hand, the electronic component of $\left< s_{x,e}\right>$ does not depend on $\beta$. The structure of electronic spin polari\-za\-tion of the zero-energy MBS is important in setting the character of the resulting magnetic profile when ave\-ra\-ging over all the degenerate MBSs occurring at the interface of the STSC junction.

\begin{figure}[t!]
\includegraphics[width=0.98\columnwidth]{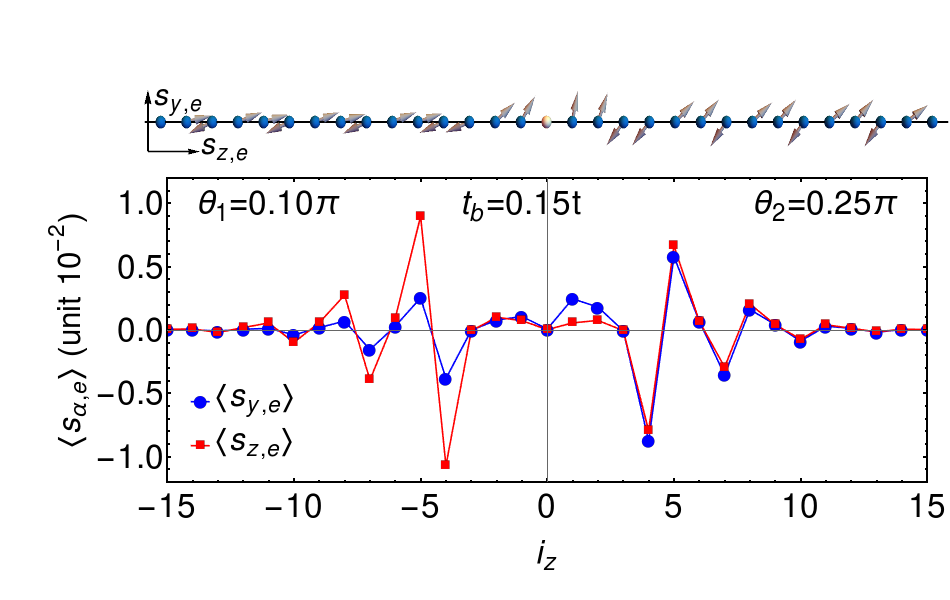}
\protect\caption{Electron component of the averaged spin-polarization over all the zero-energy modes for a representative case of the $D_{xx}$ topological phase with 4 MBSs and phase difference is $\Delta\phi_c=\pi$ and $\Delta\phi_s=0$. The parameters are: $\theta_1=0.10\pi$, $\theta_2=0.25 \pi$ and $t_b=0.15t$. In the top part of the fi\-gu\-re we schematically report on the atomically-resolved three-dimensional spin-texture of the zero-energy spin-polarization close to the interface site, with the latter being indicated as a golden point. The spin polarization lies in the $yz$ plane and exhibits a sort of antiferromagnetic pattern with an alternation of opposite pairs of neighbor spins.}
\label{fig:Figure5}
\end{figure} 

Indeed, when combining all the spin components, as shown in Fig.~\ref{fig:Figure4}(e), the resulting orientation shows a sort of regular pattern with pairs of neighbor spins pointing in the same direction, separated by a single spin which is noncollinear to them. We find that such a pattern is not sensitive to the variation of the values of the model parameters, or, changes in the applied magnetic field within the selected topological regime. Therefore, this robustness indicates a possible distinctive mark of the uncoupled interface MBSs. Such a characteristic spatial pattern is more evident in the averaged spin polarization over all the zero-energy interface MBSs. In this case, as discussed above, the $x$ component cancels out since it is independent of $\beta$, and the only nontrivial resulting orientations are in the $yz$ easy plane of the applied magnetic field. As one can see in Fig.~\ref{fig:Figure5}, the spin polarization indeed lies in the $yz$ plane and exhibits a modulated pattern with sign-changes on both sides of the junction, with a variation on the scale of the Fermi wavelength. The different orientation of the spin in the two superconductors is dictated by the misalignement of the applied field. The analysis of the MBSs in other topological regimes with two MBSs at the interface indicates that the behavior of the spin polarization for each MBSs, or averaged over the zero-energy MBSs, yields the same fingerprints in the spatial distribution and in the spin-space orientation (see Appendix~\ref{sec:Appendix}). Such a behavior is distinctive of a topological superconductor with multiple chiral-symmetry-protected boundary MBSs. 

\textcolor{black}{Finally, we also point out that the spin polarization of the standard ABSs, corresponding to in-gap bound states with nonzero energy, leads to a qualitatively different spatial profile and orientation comparably to that of the MBSs. A representative case for the $D_{xx}$ configuration is reported in Appendix~\ref{sec:Appendix}.}

\section{Networks of STSC junctions\\ and synthetic nontrivial topology}\label{sec:Networks}

So far, our analysis has been restricted to strictly 1D STSC junctions. This simplification was made under the assumption that the energy eigenvalues obtained for the actual 3D system, disperse very weakly in terms of the wave vectors defined for the remaining two dimensions. This condition is often met in real materials, e.g., in organic superconductors~\cite{Salts1,Salts2}, since such systems are construed by loosely-coupled 1D chains. Here, we aim at exploring novel to\-po\-lo\-gi\-cal phenomena that open up by consi\-de\-ring the possibility of a 2D network of 1D STSC junctions, while 3D generalizations are also possible and straightforward. However, their study will not be pursued here, since this is beyond the scope of this work. 

This investigation specifically focuses on 2D networks which extend infinitely in the second direction. Therefore, by assuming periodic boundary conditions in this direction, we introduce the wave number $q$. Consequently, the MBS operators of Sec.~\ref{sec:ABSspectra} pick up a $q$ dependence and will be now referred to as Majorana edge modes. In a similar fashion, the Majorana couplings of Sec.~\ref{sec:ABSspectra} generally depend on $q$. Since here we are interested in a qualitative discussion, we consider the limit of loosely-coupled junctions, i.e., we assume that the energy scales for the inter-junction couplings are the smallest ones and, thus, these coupling terms can be included in a perturbative fashion on top of the results found in Sec.~\ref{sec:ABSspectra}. Under these conditions, we consider that the MBS couplings discussed earlier are independent of $q$. 

The additional coupling terms relate only Majorana operators on the same side of the junction. Notably, in the absence of any inter-junction couplings the energy eigenstates of the system are dispersionless with respect to $q$, i.e., they constitute so-called flat bands. In fact, the standard, even-under-inversion and spin-conserving, inter-chain electron hopping does not affect the Majorana edge mode dispersions, at least at this level of approximation. In contrast, inter-chain p-wave pairing renders the Majorana edge modes dispersive. In this paragraph, we also restrict to p-wave pairing which only involves the $d_{x,y}$ components and set $d_z=0$, unless otherwise stated. Moreover, buil\-ding upon the results of Sec.~\ref{sec:DoubleSTSC}, we consider that either $d_x$ or $d_y$ is nonzero on a given side of a single 1D junction and proportional to $\sin k$. The inter-chain pairing contributes an additional term, which is proportional to the same or the remaining $\bm{d}$-vector component, depending on whether the pairing in $(k,q)$ space is chiral or helical~\cite{Sigrist}. In the following, we consider the $D_{xx}$, $D_{xy}$ and $\tilde{D}_{yy}$ device ground states in the presence of a ge\-ne\-ral\-ly nonzero charge-phase dif\-fe\-ren\-ce $\Delta\phi_c$, and explore the arising topological pro\-per\-ties of the Majorana-edge-mode dispersions in the synthetic $(q,\Delta\phi_c)$ space. In fact, we are interested in the possible emergence of isolated and protected gap-closing points in the ABS spectra defined in the synthetic space. See also our accompanying work discussed in Ref.~\onlinecite{PRL}. 

Before proceeding, a number of remarks are in order. Below, we do not include the $D_{yy}$ configuration in our discussion because it leads to fully-gapped ABS spectra and, thus, it is not suitable for the topological phe\-no\-me\-na we aim at investigating here. We also remark that employing a $\Delta\phi_s$ bias yields similar results. In addition, we remind the reader that the chiral and helical types of pairing considered here, do not exhaust all the pos\-si\-bi\-li\-ties of p-wave pairing. Other types exist, e.g., nematic p-wave, but are not considered here because they do not lead to a full bulk gap for zero fields. We also point out that the chiral and the helical configurations are here introduced in a non-self-consistent manner, since the cha\-rac\-ter of the present study is mainly explorative. Finally, for convenience and without loss of generality, we assume that the Majorana quasiparticles are associated with a bulk gap closing at $q=0$. This allows us to restrict our analysis to small $q$. Note that similar results can be obtained if the relevant bulk gap closing occurs at $q=\pi$.

\subsection{$\mathbf{D_{xx}}$ ground state}

Given the results of Sec.~\ref{sec:ABSspectra}, and the assumptions mentioned above, the Majorana couplings for $\Delta\phi_s=0$, connecting Majoranas across the same junction, read:  
\bea
{\cal H}_{\rm MF}^{\rm intra}(q)&=&\sum_{\nu=\pm1}t_{\nu}\cos\left[\frac{\Delta\phi_c+(1+\Pi_1\Pi_2)\pi/2}{2}\right]\times\no\\
&&i\gamma_{1;\Pi_1;\nu}(-q)\gamma_{2;\Pi_2;\nu}(q)\,.
\eea

\noi The above extension of the earlier-found formula is supplemented with the inter-junction couplings which depend on the nature of the inter-chain pairing. For chiral and helical pairing on a given side of the junction, we have $\bm{d}(k,q)=(id\sin k+d'\sin q,0,0)$ and $\bm{d}(k,q)=(id\sin k,id'\sin q,0)$, respectively, with $d'\ll d$. 

If the inter-chain pairings on both sides of the junctions are chiral, we find the additional Majorana couplings:
\bea
{\cal H}_{\rm MF}^{\rm inter}(q)=\frac{1}{2}\sum_{s=1,2}\Pi_s\sum_{\nu=\pm1}vq\gamma_{s;\Pi_s;\nu}(-q)\gamma_{s;\Pi_s;\nu}(q),\qquad
\eea

\noi where we introduced the velocity $v$, by assuming that both sides of the junction feel exactly the same inter-chain pairing. When the two sides of the junction are decoupled, the above terms induce a nonzero slope for the otherwise Majorana flat bands. However, the couplings between Majoranas across the junction allow for more complex dispersions. The structure of the couplings allows us to focus on a given $\nu$. At this stage one distinguishes two cases, i.e., whether the MBS on the two sides of the junction are dictated by the same ($\Pi_1=\Pi_2=\Pi$) or opposite ($\Pi_1=-\Pi_2=\Pi$) chirality. In the former case, the ABS energy spectrum becomes:
\bea
\epsilon_{\nu}(q,\Delta\phi_c)=\Pi vq\pm t_{\nu}\sin(\Delta\phi_c/2) 
\eea

\noi and in the latter, it takes the form:
\bea
\epsilon_{\nu}(q,\Delta\phi_c)=\pm\sqrt{(vq)^2+t_{\nu}^2\cos^2(\Delta\phi_c/2)}\,. 
\eea

\noi Evidently, only in the second situation we obtain isolated nodal points in the ABS spectra defined in the synthetic $(q,\Delta\phi_c)$ space. In this case, we find a single node located at $(q,\Delta\phi_c)=(0,\pi)$, for both $\nu=\pm1$. 

Identical results are found if the inter-chain pairing is of the helical type on both sides of the junction, with the only difference that the signs of the inter-junction couplings additionally depend on the sign of the quantum number $\nu$. In fact, we find that the Majorana couplings for the two Majorana edge modes on a given side of the junctions have opposite signs. Note, that, while this is natural for a helical p-wave STSC in the absence of magnetic fields, it is not generally expected when time-reversal symmetry and Kramers degeneracy are broken. Nevertheless, we find that this feature also persists in the weak-field limit examined here.

We now conclude with the mixed case, in which, one side of the junction is dictated by chiral pairing, and the other by helical. In this situation, we find that the Majorana edge modes of only a single $\nu$ species become coupled, and lead to a protected nodal point in the spectrum, independently of the relative sign of $\Pi_1$ and $\Pi_2$. At the same time, the two branches of the remaining species only become shifted and, thus, persist in the low-energy sector. In all the above cases, we have either found two nodes located at the same point of the synthetic space or a single node coexisting with other ungapped Majorana edge modes. Therefore, in both cases the sought-after nodal points are fragile, and all these situations correspond in practice to topologically-trivial scenarios. 

\subsection{$\mathbf{D_{xy}}$ ground state}

In this configuration, the MBSs on the right (without loss of generality) hand side of the junction are already coupled and energetically split. Considering the limit in which this splitting constitutes the largest energy scale of all the Majorana couplings, the nontrivial topology is associated with the MBSs on the left hand side of the junction. In the presence of inter-chain pairing, the effective low-energy Hamiltonian describing these two types of Majorana quasiparticles has the form:
\bea
{\cal H}_{\rm low-en}(q)&\approx&
\frac{t_+t_-}{2E_{h_{2y}}}\sin\Delta\phi_c\ph i\gamma_{1;\Pi_1;+1}(-q)\gamma_{1;\Pi_1;-1}(q)\no\\
&+&\frac{1}{2}\sum_{\nu=\pm1}\lambda_{\Pi_1,\nu} vq\gamma_{1;\Pi_1;\nu}(-q)\gamma_{1;\Pi_1;\nu}(q)\,,\qquad
\eea

\noi where $\lambda_{\Pi_1,\nu}=\lambda_{\Pi_1,-\nu}$ ($\lambda_{\Pi_1,\nu}=-\lambda_{\Pi_1,-\nu}$) for chiral (helical) inter-chain pairing on the left side of the junction. Based on the discussion of the previous paragraph, we find that only the helical pairing can lead to a topologically nontrivial nodal bandstructure. The nodes are obtained for $q=0$ and $\sin\Delta\phi_c=0\Rightarrow\Delta\phi_c=0,\pi$. These nodal points are separated in synthetic space and, therefore, are topologically protected. Interestingly, this result does not depend on the type of \textit{inter-chain} pairing emerging on the right hand side of the junction.

At this point, we briefly comment on the effects of a nonzero $d_z$, in connection with the fermion-parity pum\-ping discussed in Ref.~\onlinecite{PRL}. As shown there, the pumping occurs due to the adiabatic time dependence of a pa\-ra\-me\-ter $\theta$, which controls the composition of the Majorana couplings. For this effect to take place in the systems exa\-mi\-ned here, a nonzero $d_z$ component is required. To demonstrate this, we here assume that the STSC on the left hand side exhibits the topologically-relevant helical inter-junction pairing. Moreover, we consider that the $d_z$ component to be added is also consistent with this type of pairing. A suitable vector satisfying these constraints has the form $\bm{d}(k,q)=(id\sin k,id'\sin q,id''\sin q)$, where $d''\ll d$. The inclusion of the $d_z$ component yields the additional Majorana coupling of the form:
\bea
{\cal H}_{\rm low-en}^{d_z}(q)\propto q\sum_{\nu=\pm1}\gamma_{1;\Pi_1;\nu}(-q)\gamma_{1;\Pi_1;-\nu}(q)\,.
\eea

\noi By employing the parametrization $d'=d_{yz}\cos\theta$ and $d''=d_{yz}\sin\theta$, with $d_{yz}=\sqrt{(d')^2+(d'')^2}$, we obtain a mapping to the phenomenologically considered pum\-ping Hamiltonian of Ref.~\onlinecite{PRL}. While $\theta$ can be gauged away when it is a mere constant, this is not possible when it becomes time dependent, in which case, it further leads to the announced fermion-parity pumping effects. To sweep $\theta$ in time, one is required to rotate $\bm{d}$. This is in principle achievable by adiabatically changing the orientation of the magnetic field. Further examination of this pos\-si\-bi\-li\-ty goes beyond the scope of this work, and would require exploring the self-consistent solution of $\bm{d}$ for magnetic fields not necessarily constrained in the $yz$ spin plane.

\textcolor{black}{Concluding this section, we remind the reader that the above considerations have neglected the orbital coupling to the external magnetic fields. As pointed out in Sec.~\ref{sec:ABSspectra}, this coupling is absent for stricly 1D junctions. However, in the quasi 1D geometries of interest the orbital coupling is ge\-ne\-ral\-ly present. As one can infer from Fig.~\ref{fig:Figure1}, the orbital coupling is present (absent) for networks of chains stacked along the $x$ ($y$) direction. For a stac\-king in the $x$ direction, and thus electron motion confined to the $xz$ plane, the relevant magnetic-field component is $B_y$ and can be viewed as a result of a vector potential with a nonzero $z$ component of the form $A_z(x)=-xB_y$. The freedom to pick this Landau gauge further enables us to introduce the orbital coupling to the magnetic field by means of an effective shift of $\Delta\phi_c$ which is proportional to $xB_y\ell$. The lengthscale $\ell$ depends on the magnetic-field screening length of the p-wave SCs involved. Deep in the Meissner phase, the magnetic field survives only in the close vicinity of the junction's interface.} 

\textcolor{black}{Taking into account that there exist two nodal points in the synthetic $(q,\Delta\phi_c)$ space, which yield two massless Dirac Hamiltonians for the ABSs, one finds that the addition of the $x$ dependence of $\Delta\phi_c$ will lead to the emergence of relativistic Landau levels similar to the case of graphene~\cite{Graphene}. Therefore, one expects the appea\-ran\-ce of a zero-energy Landau level and the emergence of an anomalous response, which are linked to the fermion-parity pumping effects discussed above and in Ref.~\onlinecite{PRL}.}

\subsection{$\mathbf{\tilde{D}_{yy}}$ ground state}

Similar arguments to the ones presented in the paragraph above, allow us to conclude that the topologically-relevant case is when helical pairing dictates the side of the junction which feels the magnetic field oriented along the $z$ direction. Nevertheless, in contrast to the $D_{xy}$ case, here we find a second-order node located at $(q,\Delta\phi_c)=(0,\pi)$. The order here reflects that an expansion of the energy dispersion about this node is quadratic in $\Delta\phi_c$. However, a second-order node can be viewed as two merged linear order nodes. The latter implies that this node is unstable and the arising topological properties trivial.

\section{Conclusions}\label{sec:Conclusions}

\begin{table*}[t!]
\begin{center}
\begin{tabular}{|c|c|c|c|c|c|}
\hline
\begin{tabular}{c}
Superconducting \\ Ground State
\end{tabular}
& 
\begin{tabular}{c}
Magnetic Field \\ Configuration
\end{tabular}
& 
\begin{tabular}{c}
Electrically-Driven\\  Transition
\end{tabular}
& 
\begin{tabular}{c}
Interfacial Majorana \\Bound States (MBSs)
\end{tabular}
&
\begin{tabular}{c}
Electronic Spin \\ Polarization \\ of a Single MBS
\end{tabular}
&
\begin{tabular}{c}
Averaged \\ Electronic Spin \\ Polarization
\end{tabular}
\\
\hline
\begin{tabular}{c}
$D_{xx}$ \\ SC$_1$ [$d_x$] - SC$_2$ [$d_x$]
\end{tabular}
& $\theta_{1,2} \in [\alpha , \pi-\alpha]$  & 
\begin{tabular}{c}
No Phase \\ Transition
\end{tabular}
&
\begin{tabular}{c}
4 MBSs for \\  $(\Delta\phi_c,\Delta\phi_s)=(\pi,0)$
\end{tabular}
&   
\begin{tabular}{c}
$s_{\alpha,e}\neq 0$ \\ $\alpha=x,y,z$
\end{tabular} 
&
\begin{tabular}{c}
$ \langle s_{x,e} \rangle =0$\\ $ \langle s_{(y,z),e} \rangle \neq 0$
\end{tabular}
\\
\hline
\begin{tabular}{c}
$D_{xy}$ \\ SC$_1$ [$d_x$] - SC$_2$ [$d_y$]
\end{tabular}
&
\begin{tabular}{c}
 $\theta_1 \in [\alpha,\pi-\alpha]$ and \\$\theta_2 \in [0,\alpha]$ or \\
 $\theta_2 \in [\pi-\alpha,\pi]$
\end{tabular}
&  $D_{xy} \to D_{xx}$  & 
\begin{tabular}{c}
2  MBSs for \\ $(\Delta\phi_c,\Delta\phi_s)=(0,0)$ \\ $(\Delta\phi_c,\Delta\phi_s)=(\pi,0)$ \\
$(\Delta\phi_c,\Delta\phi_s)=(0,\pi)$
\end{tabular}
&  \begin{tabular}{c}
$s_{\alpha,e}\neq 0$ \\ $\alpha=x,y,z$
\end{tabular} 
&
\begin{tabular}{c}
$ \langle s_{x,e} \rangle =0$\\ $ \langle s_{(y,z),e} \rangle \neq 0$
\end{tabular}\\
\hline
\begin{tabular}{c}
$D_{yy}$ \\ SC$_1$ [$d_y$] - SC$_2$ [$d_y$]
\end{tabular}
&
\begin{tabular}{c}
$\theta_i \in [0,\alpha]$ \\ $\theta_j \in [\pi-\alpha,\pi]$ \\ $i,j=1,2$ 
\end{tabular}   
& 
\begin{tabular}{c}
$D_{yy} \to D_{xx}$ or \\$D_{yy} \to D_{xy} \to D_{xx}$  
\end{tabular}
&
\begin{tabular}{c}
4  MBSs for \\$(\Delta\phi_c,\Delta\phi_s)=(0,\pi)$
\end{tabular}
  &
\begin{tabular}{c}
$s_{\alpha,e}\neq 0$ \\ $\alpha=x,y,z$
\end{tabular} 
&
\begin{tabular}{c}
$ \langle s_{x,e} \rangle =0$\\ $ \langle s_{(y,z),e} \rangle \neq 0$
\end{tabular}\\
\hline
\begin{tabular}{c}
$\tilde{D}_{yy}$ \\ SC$_1$ [$d_y$] - SC$_2$ [$d_y$]
\end{tabular}
&
\begin{tabular}{c}
 $\theta_i=0,\pi$ and \\ $\theta_j \in [0,\alpha]$ or \\ $\theta_j \in [\pi-\alpha,\pi]$ \\
$ i,j=1,2 \;(i\neq j)$
\end{tabular}
 & $\tilde{D}_{yy} \to D_{xx}$ &
\begin{tabular}{c}
 2  MBSs for \\  $(\Delta\phi_c,\Delta\phi_s)=(\pi,0)$ \\4  MBSs for \\$(\Delta\phi_c,\Delta\phi_s)=(0,\pi)$
\end{tabular}
&  \begin{tabular}{c}
$s_{x,e}=0$ \\$s_{y,e}\neq 0 $ \\$s_{z,e}\neq 0 $
\end{tabular} 
&
\begin{tabular}{c}
$ \langle s_{x,e} \rangle =0$\\ $ \langle s_{y,e} \rangle =0$\\$ \langle s_{z,e} \rangle \neq 0$
\end{tabular}\\
\hline
\end{tabular}
\caption{\color{black}{Summary of our main results for the investigated heterostructure for the various configurations of the $\bm{d}$ vector and Zeeman/exchange field $\bm{h}$. The spin exchange fields $\bm{h}_{1,2}$ lie in the $yz$ plane and act on the two sides of the junction. Their orientation is set by the angles $\theta_{1,2}$, respectively. In the first column, we report on the superconducting ground state corresponding to the magnetic configurations given in the second column. Here, the critical angle $\alpha$ depends on the amplitude of the Zeeman/exchange field $h$ and on the charge-transfer strength $t_b$. For instance, for the parameters $h=0.05t$, $t_b=0.05t$ and $\mu=1.0t$ one obtains $\alpha\simeq0.15\pi$. The corresponding possible transitions that can be driven by electrically gating the weak link (i.e. tuning $t_b$) are indicated in the third column. The details concerning the first three columns are presented in Sec.~\ref{sec:DoubleSTSC} and Fig.~\ref{fig:Figure2}. In the last three columns, we provide the essential aspects related with the topological behavior of the p-wave Josephson junction. In particular, we present the number of MBSs harbored at the interface when a charge- or spin-phase difference across the junction is applied (see also Sec.~\ref{sec:ABSspectra}). Finally, the electron component of the spin-polarization of a single interfacial MBS and the averaged spin character over all the zero-energy bound states are summarized in the last two columns. The detailed analysis is reported in Sec.~\ref{sec:SpinPolarization} and Appendix~\ref{sec:Appendix}.}}
\label{Table:T1}
\end{center}
\end{table*}

\textcolor{black}{Before starting the discussion, we first summarize our overall results for the investigated spin-triplet junction in Table~\ref{Table:T1}. There, we report the key outcomes of the analysis concerning: i) the type of the ground state of each spin-triplet superconductor, ii)  the angle configuration of the applied magnetic fields across the 1D junction,  iii)~the sequence of superconducting ground-state transitions which are induced upon tuning the strength of the charge transfer, iv) the charge- and spin-phase values, for which, interface Majorana bound states become accessible, and v) the electronic spin polarization generated by the Majorana bound states.}

\textcolor{black}{In more detail, in this work, we} show that the vectorial nature of the superconducting order parameter, which fully encodes the information of the spin-triplet structure of the Cooper pairs, is a prominent and essential handle to design the topological and transport properties of superconducting he\-te\-ro\-structu\-res. While the odd-parity character of the superconducting order parameter is well known to be a prerequisite for ha\-ving topologically-protected bound states, we demonstrate how the presence of spin-active degrees of freedom can lead to new symmetry-enriched physical scenarios in the context of topological superconductors. 
 
From the point of view of the control physical me\-cha\-nisms, we single out unique magnetoelectric means to drive topological transitions without any gap closing, by fully exploiting the nontrivial connection between bulk and boundary states when different configurations of the superconducting order parameter compete and are close in energy. 
The here-proposed effects clearly indicate that spin-dependent and boundary-driven reconstruction can be realized in superconductors with spin-active degrees of freedom and can remarkably lead to a {\it local} control of the topological properties of hybrid superconducting he\-te\-ro\-struc\-tu\-res. 

Futhermore, the performed analysis allows us to find two distinct physical and functional regimes for the topological Josephson junction. When gating the interface into a configuration with a large tunnel-barrier amplitude and for magnetic-field orientations mostly perpendicular to the $\bm{d}$-vector plane, the system exhibits conventional Andreev-bound-state spectra which can be converted into Majorana bound states only by applying a $\pi$ spin-phase difference across the junction. In the other regions of the phase diagram one can generally bring the system to a regime of large tunability of the Andreev spectra by directly manipulating the configurations with two or four Majorana bound states through either a va\-ria\-tion of the magnetic field orientation in one or both sides of the superconductors or by modifying the interface transparency by electrically gating the weak link as well as by suitably adjusting the phase drop across the junction. In this regime, we have at hands an extraordinary flexibility in the control of the hybrid junction as it can be magnetoelectrically guided to switch between different configurations with multiple Majoranas at the interface. 

The distinct mark of the behavior of the topological Josephson junction is reflected in the wide range of regimes of Andreev spectra that can be obtained by both va\-rying the magnetic-field orientation and the strength of the interface transparency. The versatility of the spin- and electric-active topological junction is also demonstrated by the rich variety of Andreev spectra which can be designed when a charge- and/or spin-phase twists the superconducting order parameters across the interface. The inequivalence of the Andreev spectra upon the application of a spin- or a charge-phase difference also implies a sort of spin-charge separation with the possibility of an independent tunability of the superconducting transport properties of the junction in the charge and spin sector. Even more importantly, the novel Andreev dispersions found here, can be also employed to engineer synthetic nontrivial topology when considering 2D networks of such 1D junctions. Our findings reveal that the case of a helical p-wave superconductor in the $D_{xy}$ con\-fi\-gu\-ra\-tion constitutes the most prominent scenario towards realizing this type of phenomena.

\textcolor{black}{One relevant issue concerns the type of magnetic means and superconducting materials that can bring the proposed heterostructure into realization. Regarding the magnetic control, there exist various ways which can be employed to locally modify the Zeeman field. The first possibility relies on applying a rotating magnetic field generated by a so-called vector magnet, which can provide a 3D control on the field orientation with an angle resolution finer than 0.1 degrees. At the same time, one is required to implement a suitable geometric junction design that leads to misaligned ${\bm d}$ vectors across the junction, already at zero external magnetic fields. In this manner, rotating the vector magnet allows accessing different regimes of the topological phase diagram.}

\textcolor{black}{Alternative solutions involve local magnetic or electric controls that become accessible by employing ferroic materials. For instance, by means of ferromagnets in pro\-xi\-mi\-ty with the superconductor, one can induce an effective spin exchange on the spin-triplet pairs. The use of ferromagnetic materials can further expand the control of the local exchange field, since imposing a superconducting phase difference across a magnetic domain wall can also define a Josephson junction. Ideal ma\-te\-rials for the realization of such a heterostructure are the heavy-fermion ferromagnetic superconductors, e.g. UGe, URhGe, UGeCo, where ferromagnetism and spin-triplet superconductivity generally coexist in a large region of the phase diagram~\cite{Aoki19}. A local control on the orientation of the magnetization field can be also achieved indirectly, by applying suitable electric fields. This becomes possible by virtue of the magnetoelectric coupling encountered in multiferroics~\cite{Matsukura15}. The local control of an electric field at the nanometer scale is advantageous compared to the direct manipulation of the magnetization, and is feasible through gating or by employing electrical tips.}

\textcolor{black}{On the side of materials, we point out that there exists a long list of quasi 1D systems that have already shown experimental fingerprints of spin-triplet superconducting pairing and appear as prominent building blocks for fabricating the proposed hybrid devices. For instance, this is the case for the organic Bechgaard salts~\cite{Salts1,Salts2}, the purple molybdenum bronze Li$_{0.9}$Mo$_6$O$_{17}$~\cite{purple}, and more recently the Cr-based pnictide superconductors~\cite{Bao15,Cuono19}. Moreover, we remark that the design of spinful triplet superconductivity can also rely on harnessing the orbital degrees of freedom, as it has been demonstrated for 2D electron gases~\cite{Fukaya18} and multi-orbital optical-lattice systems exhibiting superfluidity~\cite{Panahi12}.}

Apart from the large versatility in the generation and manipulation of different topological configurations, the examined heterostructure has distinct magnetic properties when probing the spatially-resolved spin polarization of the Majorana and Andreev bound states. In the topological regime with more than one Majorana bound state at the edge, each state is marked by an Ising-type spin-polarization, which is always pointing along a direction perpendicular to the easy plane of the applied magnetic field. Furthermore, local tunnel spectroscopy can give further access to the magnetic properties of the Majorana bound states, thus unveiling a richer internal structure corresponding to the electron-hole components of the Majorana wave function. Indeed, each Majorana bound state is marked by a spin pattern with a 3D spin texture. Then, the chiral symmetry makes the averaged spin polarization at zero voltage to yield a spatially mo\-du\-la\-ted profile at the atomic scale that is tied to stay in the easy plane of the applied magnetic field. Such predictions set remarkable symmetry constraints for the topological junctions with multiple Majorana bound states per edge and are significantly relevant for their unambiguous detection.

\textcolor{black}{Finally, it is also worth discussing to which extent the conclusions of our analysis are related to the presence of atomic spin-orbit coupling. Indeed, for the materials of interest, it is the combination of the atomic spin-orbit coupling and the crystalline-field potential which ty\-pi\-cal\-ly sets the magnetic anisotropy. This is by assuming that there exist multi-orbital degrees of freedom close to the Fermi level, as well as that both time-reversal and inversion symmetries are present. The magnetic anisotropy terms tend to pin the ${\bm{d}}$ vector, and introduce an easy axis or plane. For instance, at lowest order in the local total spin momentum, one finds the magnetic-anisotropy contribution to the Hamiltonian $H_{\rm mag}=a_x S_x^2+ a_y S_y^2+a_z S_z^2$, with $a_i$ the anisotropic coefficients along the crystal symmetry axes. In our analysis, we are already effectively including the consequence of magnetic anisotropy, since we are considering that the interactions driving the spin-triplet pairing lead to an easy plane anisotropy with preferred $(d_x,d_y)$ components while $d_z$ is vanishing. Such assumption is compatible with the low-dimensional cha\-rac\-ter of the superconductor and the considered crystal symmetry. Hence, our study is generally applicable to the phy\-si\-cal case of an easy plane magnetic anisotropy which set the $\bm{d}$ vector to lie in a plane. Furthermore, even for a magnetic-anisotropy coupling that favors an easy axis spin orientation, the presented results are still valid as long as the free-energy difference between magnetic configurations developing along the easy and hard axes, is smaller than the applied magnetic field.}

\appendix

\section{Spin polarization of bound states}
\label{sec:Appendix}

\subsection{ABS- and MBS-induced spin-polarization for the ${\bf D_{xx}}$ configuration}
\label{sec:Appendix1}

\begin{figure}[t!]
\includegraphics[width=0.48\columnwidth]{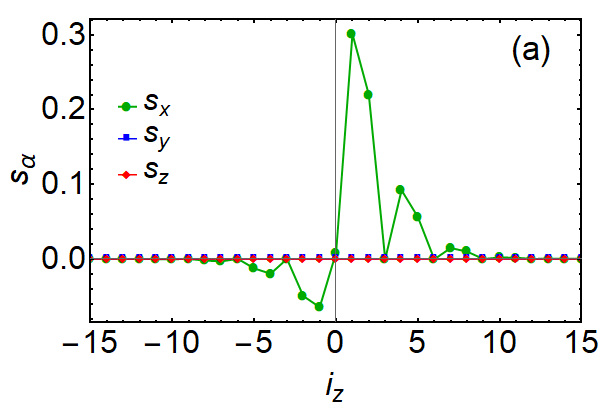}
\includegraphics[width=0.5\columnwidth]{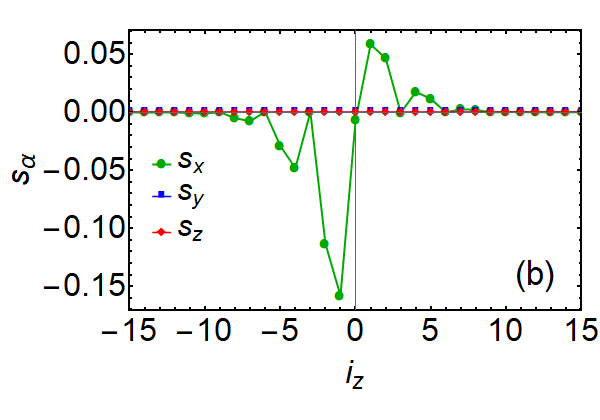}\\
\includegraphics[width=0.49\columnwidth]{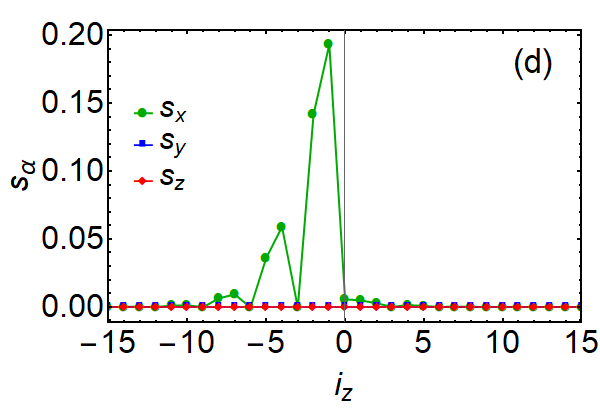}
\includegraphics[width=0.49\columnwidth]{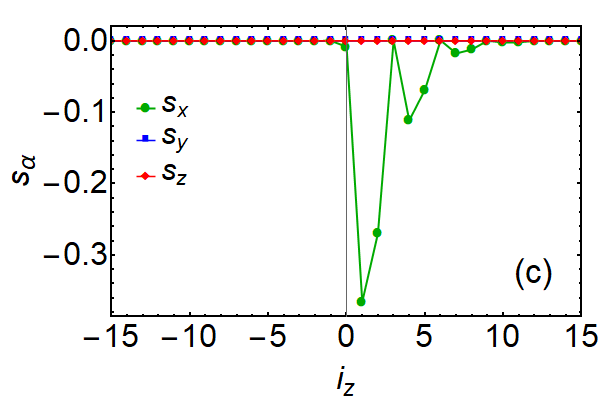}
\protect\caption{\textcolor{black}{Spin-polarization for each one of the interfacial MBS, for a representative case of the $D_{xx}$ ground state. Parameter values: $\theta_1=0.10\pi$, $\theta_2=0.25\pi$, $t_b=0.15t$ and $\Delta\phi_{c}=\pi$.}}
\label{fig:Figure6}
\end{figure}

\textcolor{black}{To complete the analysis of Sec.~\ref{sec:SpinPolarization}, we show here the spin-polarization (i.e. including \textit{both} electron and hole contributions) for the $D_{xx}$ configuration. As discussed in the main text, the spin-polarization of a single MBS yields a nonzero spin polarisation only for the $x$ component (see Fig.~\ref{fig:Figure6}), while the average over all MBS is zero.}  

\begin{figure}[t!]
\includegraphics[width=0.48\columnwidth]{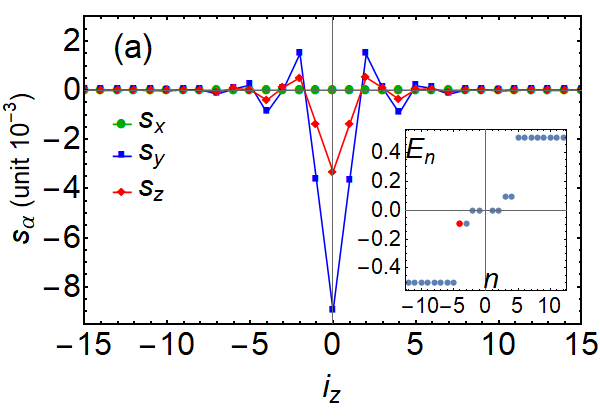}
\includegraphics[width=0.48\columnwidth]{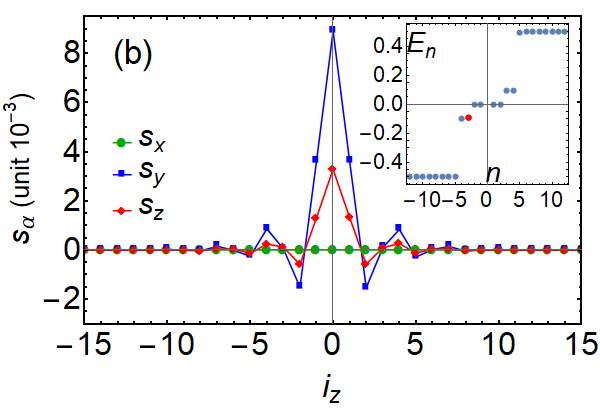}
\\
\includegraphics[width=0.49\columnwidth]{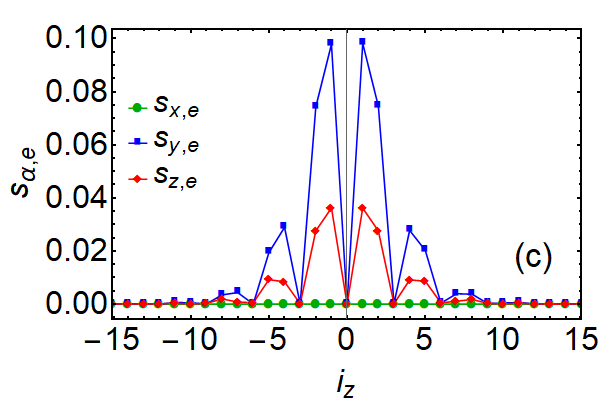}
\includegraphics[width=0.49\columnwidth]{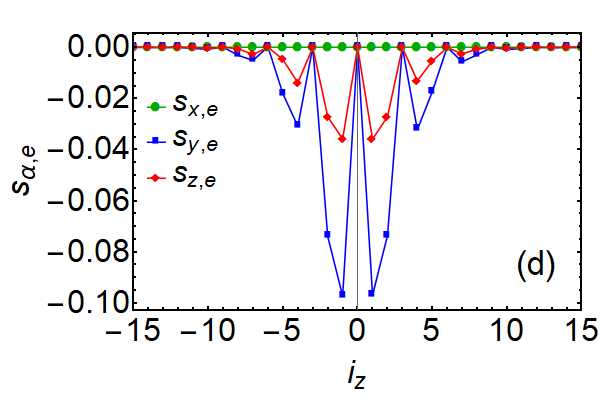}
\protect\caption{\textcolor{black}{(a)-(b) Spin-polarization (including both electron and hole contributions) of the first two interfacial ABS (at negative energy), for a representative case of the $D_{xx}$ ground state. In the inset, the low energy levels are plotted and the energy of the considered ABS is marked with a red dot. Here, the ABS energies are multiplied by a factor of 10 for clarity of visualization of the energy gap. The four zero energy states correspond to the MBS at the external edges of the junction. (c)-(d) Electronic spin-polarization of the ABSs considered in (a) and (b). Parameter values: $\theta_1=0.10\pi$, $\theta_2=0.25\pi$, $t_b=0.15t$ and $\Delta\phi_{c,s}=0$}.}
\label{fig:Figure7}
\end{figure}

\textcolor{black}{Next, we consider the behavior of the spin-polarization for the standard Andreev bound states that arise due to the hybridization of the Majorana bound states at the interface. These results are reported in Fig.~\ref{fig:Figure7}. The spin-polarization, obtained after summing up the electron and hole contributions, is vanishing for the $x$ orientation, in contrast to the MBS case, while it has a nonvanishing spatial profile for the other two directions (see Fig.~\ref{fig:Figure7}(a)-(b)). Re\-mar\-ka\-bly, the spin-polarization has a symmetric distribution across the interface. A similar behavior is also observed for the electronic component of the ABS spin-polarization (see Fig.~\ref{fig:Figure7}(c)-(d)).}

\subsection{MBS-induced electronic spin-polarization for ${\bf D_{xy}}$ and ${\bf \tilde{D}_{yy}}$ configurations}
\label{sec:Appendix2}

\begin{figure*}[t!]
\includegraphics[width=0.92\columnwidth]{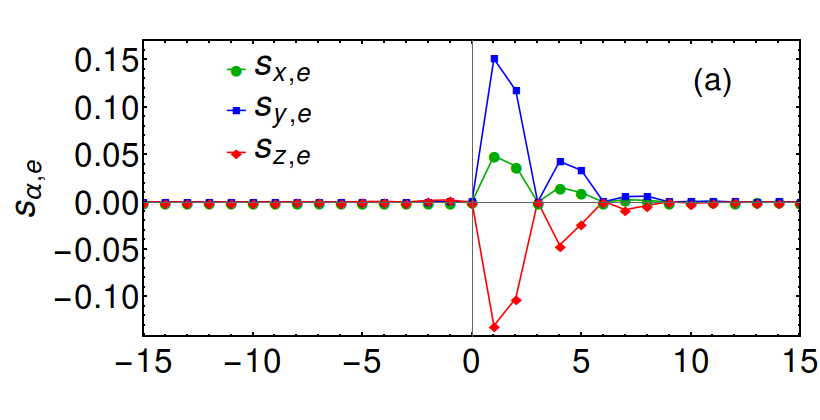}\hspace{0.8cm}
\includegraphics[width=0.92\columnwidth]{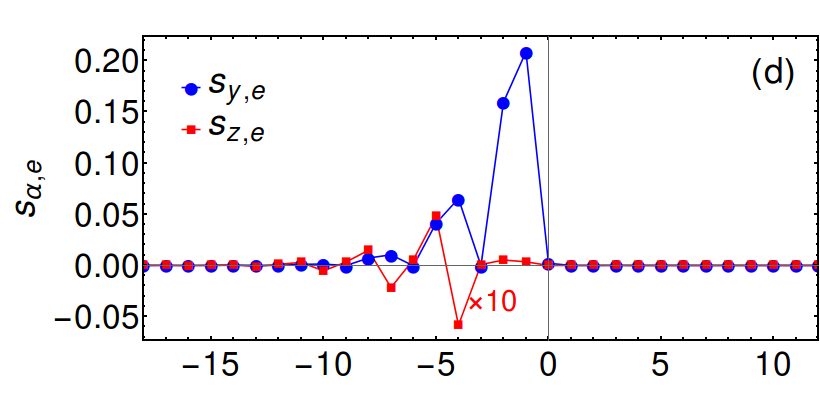}\vspace{-0.28cm}\\
\includegraphics[width=0.92\columnwidth]{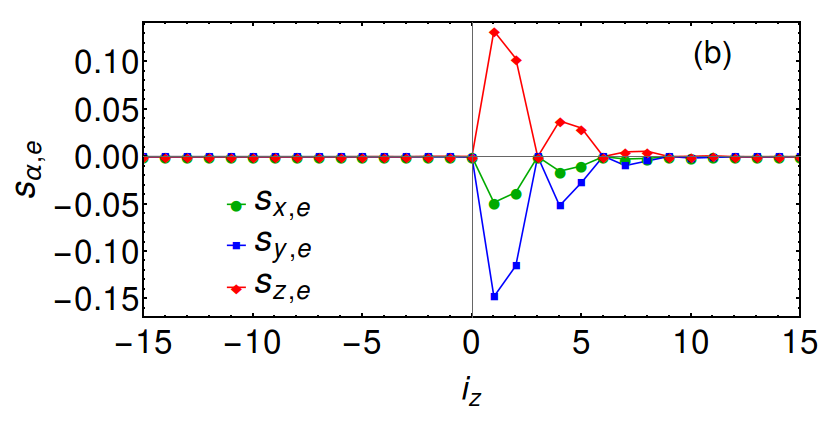}\hspace{0.8cm}
\includegraphics[width=0.92\columnwidth]{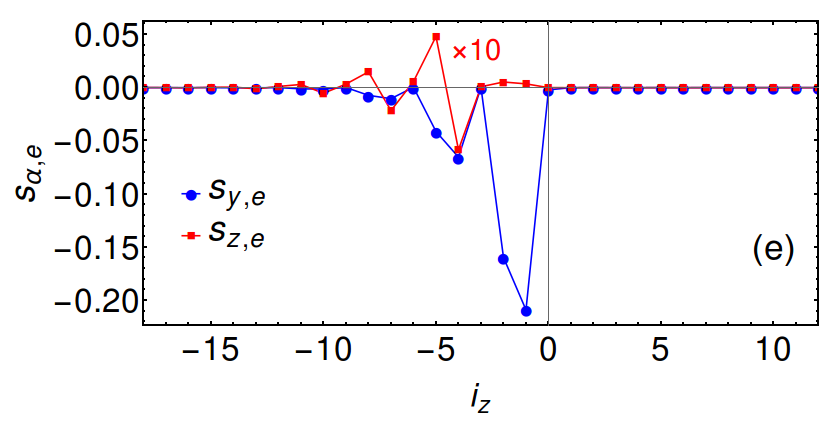}\\
\includegraphics[width=0.98\columnwidth]{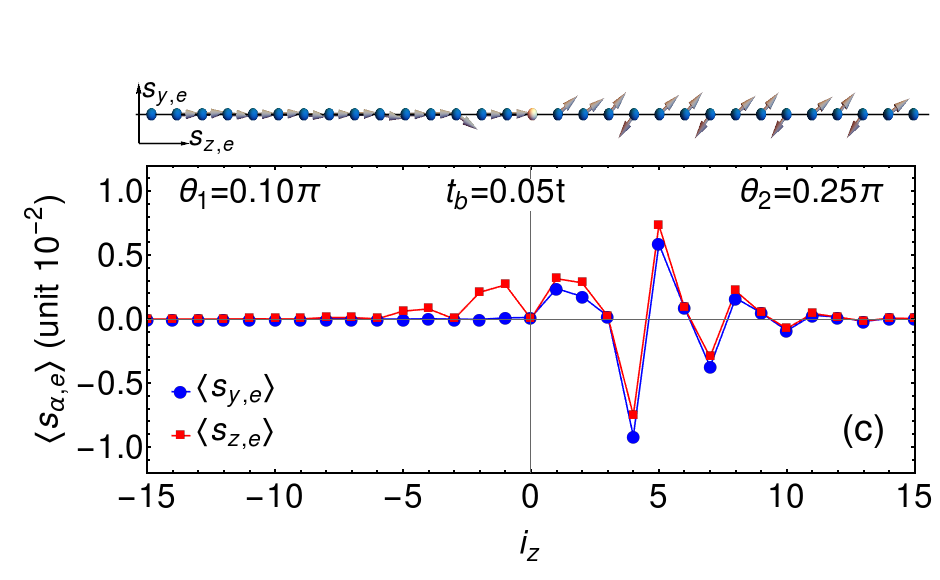}
\includegraphics[width=0.98\columnwidth]{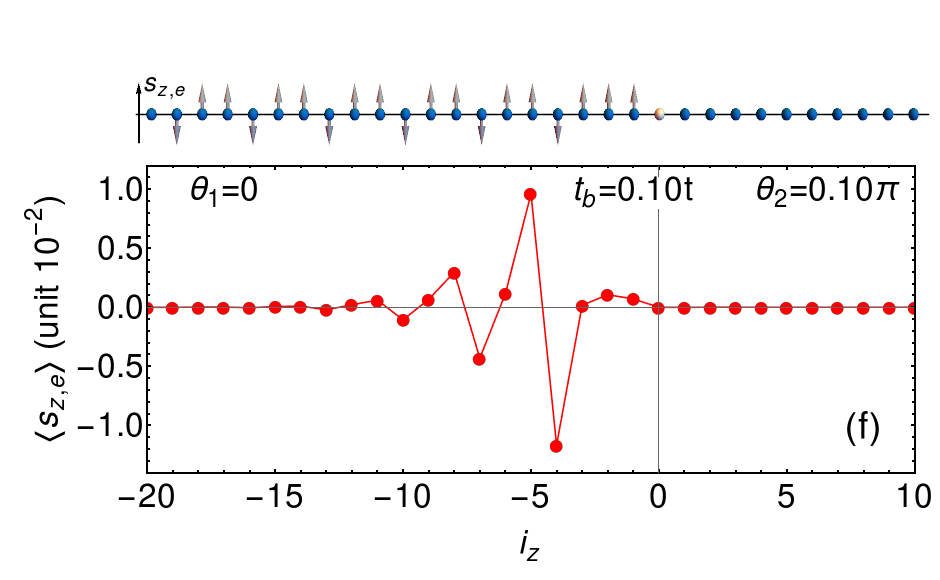}
\caption{(a)-(b) Electron component of the spin polarization of each MBS at the interface in a representative case of the $D_{xy}$ phase. In this region there are 2 MBS at the interface when the phase difference is $\Delta\phi_{c,s}=0$ or $\pi$. The parameters are: $\theta_1=0.10\pi$, $\theta_2=0.25 \pi$,  $t_b=0.05t$ and $\Delta\phi_{c,s}=0$. (c) Averaged spin polarization over the zero-energy modes at the interface as in (a)-(b). The averaged $\langle s_{x,e}\rangle$ component is zero. (d)-(e) Electron component of the spin polarization of each MBS at the interface for representative case of the $\tilde{D}_{yy}$ phase, with 2 MBSs at a phase difference $\Delta\phi_c=\pi$ and $\Delta\phi_s=0$. The parameters are: $\theta_1=0$, $\theta_2=0.1 \pi$ and $t_b=0.1t$. The $s_{x,e}$ component of each MBS is zero (not shown). (f) Averaged spin polarization over the zero-energy modes at the interface as in (d)-(e). The only nonzero averaged component in this case is $\langle s_{z,e}\rangle$.}
\label{fig:Figure8}
\end{figure*}

To \textcolor{black}{complement} the study of Sec.~\ref{sec:SpinPolarization}, here we present the evolution of the electronic Majorana-induced spin-polarization for the $D_{xy}$ and $\tilde{D}_{yy}$ configurations of the superconducting heterostructure. The results are reported in Fig.~\ref{fig:Figure8}. Similar to the $D_{xx}$ state, the electron spin polarization for each bound state has a 3D profile in spin space, and it is spatially asymmetric with respect to the interface because the Majorana states are mainly located on the topologically non-trivial side of the junction (See Figs.~\ref{fig:Figure8}(a),(b),(d)~and~(e)). The average amplitude over the zero-energy degenerate configurations indicates a $yz$ planar spin orientation for the $D_{xy}$ case, with the spin component perpendicular to the magnetic field direction that is identically zero. This is shown in Fig.~\ref{fig:Figure8}(c). On the other hand, the averaged spin profile for the $\tilde{D}_{yy}$ is Ising-type with only a non-vanishing $z$-axis projection, which shows modulations in sign and amplitude near the interface, see Fig.~\ref{fig:Figure8}(f). 

\end{document}